\documentclass{aa}
\usepackage[varg]{txfonts}
\usepackage{graphicx}

\begin{document}
\title{Correlation of gas dynamics and dust in the evolved filament G82.65-02.00\thanks{$Herschel$ is an ESA space observatory with science instruments provided by European-led Principal Investigator consortia and with important participation from NASA}}

\author{M. Saajasto\inst{1}
  \and M. Juvela\inst{1}$^{,}$\inst{2}
  \and K. Dobashi\inst{3} 
  \and T. Shimoikura\inst{3} 
  \and I. Ristorcelli\inst{4}$^{,}$\inst{5} 
  \and J. Montillaud\inst{2}
  \and D.J. Marshall\inst{6}
  \and J. Malinen\inst{7}$^{,}$\inst{8}
  \and V.-M. Pelkonen\inst{1}$^{,}$\inst{2}$^{,}$\inst{9}
  \and O. Fehér\inst{10}
  \and A. Rivera-Ingraham\inst{11}
  \and L. V. Toth\inst{10}
  \and L. Montier\inst{4}$^{,}$\inst{2} 
  \and J-Ph. Bernard\inst{4}$^{,}$\inst{2}
  \and T. Onishi\inst{12}  
  }


\institute{Department of Physics, P.O.Box 64, FI-00014, University of Helsinki, Finland
\and Institut UTINAM - UMR 6213 - CNRS - Univ Bourgogne Franche Comt\'e
\and Department of Astronomy and Earth Sciences, Tokyo Gakugei University, Koganei, Tokyo 184-8501, Japan
\and Université de Toulouse, UPS-OMP, IRAP, 31028 Toulouse, France
\and CNRS, IRAP, 9 Av. Colonel Roche, BP 44346, 31028 Toulouse Cedex 4, France
\and Laboratoire AIM, IRFU$/$Service d’Astrophysique - CEA$/$DSM - CNRS - Universit\'e Paris Diderot, France
\and Department of Physics, Florida State University, Tallahassee, FL 32306, USA
\and Institute of Physics I, University of Cologne, Germany
\and Finnish Centre for Astronomy with ESO, University of Turku, V\"ais\"al\"antie 20, FI-21500 PIIKKI\"O, Finland
\and Eötvös University, Department of Astronomy, Pázmány P. s. 1$/$a, 1117 Budapest, Hungary
\and European Space Astronomy Centre ESA$/$ESAC, PO Box 78, 28691 Villanueva de la Cañada, Madrid, Spain
\and University of Osaka Prefecture
}

\date{Received day month year / Accepted day month year}

\abstract{The combination of line and continuum observations can provide vital insights to the formation and fragmentation of filaments and the initial conditions for star formation. We have carried out line observations to map the kinematics of an evolved, actively star forming filament G82.65-2.00. The filament was first identified from the $Planck$ data as a region of particularly cold dust emission and was mapped at 100-500 $\mu$m as a part of $Herschel$ key program Galactic Cold Cores. The $Herschel$ observations cover the central part of the filament, corresponding to a filament length of $\sim 12$ pc at the assumed distance of 620 pc.}
{CO observations show that the filament has an intriguing velocity field with several velocity components around the filament. In this paper, we study the velocity structure in detail, to quantify possible mass accretion rate onto the filament, and study the masses of the cold cores located in the filament.}
{We have carried out line observations of several molecules, including CO isotopologues, $\rm HCO^{+}$, HCN, and CS with the Osaka 1.85 m telescope and the Nobeyama 45 m telescope. The spectral line data is used to derive velocity and column density information.}
{The observations reveal several velocity components in the field, with strongest line emission concentrated to velocity range $\sim [3,5]$  km $\rm s^{-1}$. The column density of molecular hydrogen along the filament varies from 1.0 to 2.3 $\times 10^{22}$ $\rm cm^{2}$. We have examined six cold clumps from the central part of the filament. The clumps have masses in the range $10 - 20$ $M_{\odot}$ ($\sim70M_{\odot}$ in total) and are close to or above the virial mass. Furthermore, the main filament is heavily fragmented and most of the the substructures have a mass lower than or close to the virial mass, suggesting that the filament is dispersing as a whole. Position-velocity maps of $\rm ^{12}CO$ and $\rm ^{13}CO$ lines indicate that at least one of the striations is kinematically connected to two of the clumps, potentially indicating mass accretion from the striation onto the main filament. We tentatively estimate the accretion rate to be $\dot{M}$ = $2.23 \times 10^{-6}$ $ M_{\odot} / \rm year$.}
{Our line observations have revealed two, possibly three, velocity components connected to the filament G82.65-2.00 and possibly signs of mass accretion onto the filament. The line observations combined with $Herschel$ and $WISE$ maps suggest a possible collision between two cloud components.}

\maketitle

\section{Introduction}\label{Sec1}

Filamentary structures have been found to be common features of interstellar molecular clouds. The early results from the $Herschel$ space observatory have shown the abundance of $\sim$1 parsec scale structures even in the diffuse and non-star-forming clouds \citep{Andre2010, Arzoumanian2011, Menshchikov2010, Juvela2012, RiveraIngraham2016}. Based on these observational results, it has been suggested that filamentary structures form as a result of the dissipation of large-scale turbulence and not as a result of star formation \citep{Arzoumanian2011}. On the other hand, the filamentary structures have been found to be closely connected to star formation, especially in the nearby fields where the $Herschel$ resolution is sufficient to resolve gravitationally bound cores. The close relationship between cores and filaments lead \citet{Menshchikov2010} and \citet{Konyves2010} to suggest that core formation follows a two step path. The large scale turbulence and magnetic fields generate a complex web of filamentary structures followed by gravitational fragmentation of the densest structures.

Understanding the morphology and fragmentation of filamentary structures is vital in order to understand star formation \citep[e.g.][]{Inutsuka1992, Kainulainen2016}. On the other hand, the internal motions of filaments and interactions between the filament and surrounding medium can provide insights on the evolution from cores to stars \citep{Miettinen2013}. In the L1517 and B213 clouds in the Taurus cloud complex, a number of velocity-coherent structures were identified from $\rm C^{18}O$ and $\rm N_2H^{+}$ data by \citet{Hacar2011} and \citet{Hacar2013}. The coherent nature of the structures was interpreted as evidence of physically distinct sub-filaments. The authors suggested that the structures have formed at different times and only some of them are producing dense cores. Furthermore, the filaments identified in the nearby cloud L1517 were found to be surrounded by subsonic gas \citep{Hacar2011}, suggesting that the formation of the velocity-coherent structures has occurred before fragmentation. The formation of velocity coherent substructures has been connected to the evolutionary stage of the filament. In their recent paper, \citet{Cox2016} studied the Musca filament and compared its structure with the B211/213 filament system. Both of the filaments are quite similar with 10 pc long main filament and are surrounded by a number of fainter and roughly perpendicular substructures. However, the B211/213 filament has complex internal velocity coherent structures while there is significantly less substructure in Musca. The differences led \citet{Cox2016} to suggest that the Musca cloud represents an earlier evolutionary stage.

The faint substructures, striations, seen around the filaments can be related to mass accretion from the surrounding medium on the main filament. The findings of \citet{Arzoumanian2013} and \citet{Palmeirim2013} point to a scenario where filaments continue to accrete material throughout their evolution while retaining approximate virial balance. However, for non-equilibrium filaments, the scenario between accretion rate and fragmentation might be more complicated as discussed by \citet{Clarke2016}.

The formation of large scale structures that are divided to smaller sub-filaments is also seen in simulations as discussed by \citet{Smith2016}. In their hydrodynamic turbulent cloud simulations, the sub-filaments were formed naturally and not as a result of fragmentation. Thus, core formation seems to have little or no effect on gas kinematics, suggesting that core formation through violent cloud-to-cloud collision is rare, although it is still possible \citep{Balfour2015}.

Despite the numerous observational constraints, the exact formation and destruction methods of filamentary structures are still an open question. \citet{RiveraIngraham2016, RiveraIngraham2017} studied a wide range of close-by (D$<$500 pc) filaments present in the $Herschel$ fields. A significant dispersion in the observed filament properties suggests that the evolutionary paths of filaments are highly dependent on the external conditions, such as Galactic location or main physical processes. \citet{RiveraIngraham2016, RiveraIngraham2017} do not explicitly examine the role of  magnetic fields in filament formation. The orientation of magnetic fields has been observed to be linked to column density \citep{PlanckXXXV, Cox2016}. For instance in the high latitude cloud L1642 \citep{Malinen2016} the magnetic field shows a clear transition in direction from aligned to perpendicular to the cloud elongation approximately at a column density of $N_{\rm H}$ = $1.6 \times 10^{21}$ $\rm cm^{-2}$. The findings of \citet{RiveraIngraham2016, RiveraIngraham2017} suggest that the formation and evolution of filament is driven by complex non-equilibrium processes between intrinsic and environmental conditions. Thus, understanding the differences in the primordial local conditions are vital in order to determine the structure and evolution of the clouds as well as star formation therein.

\begin{figure*}
\centering
\includegraphics[width=17cm]{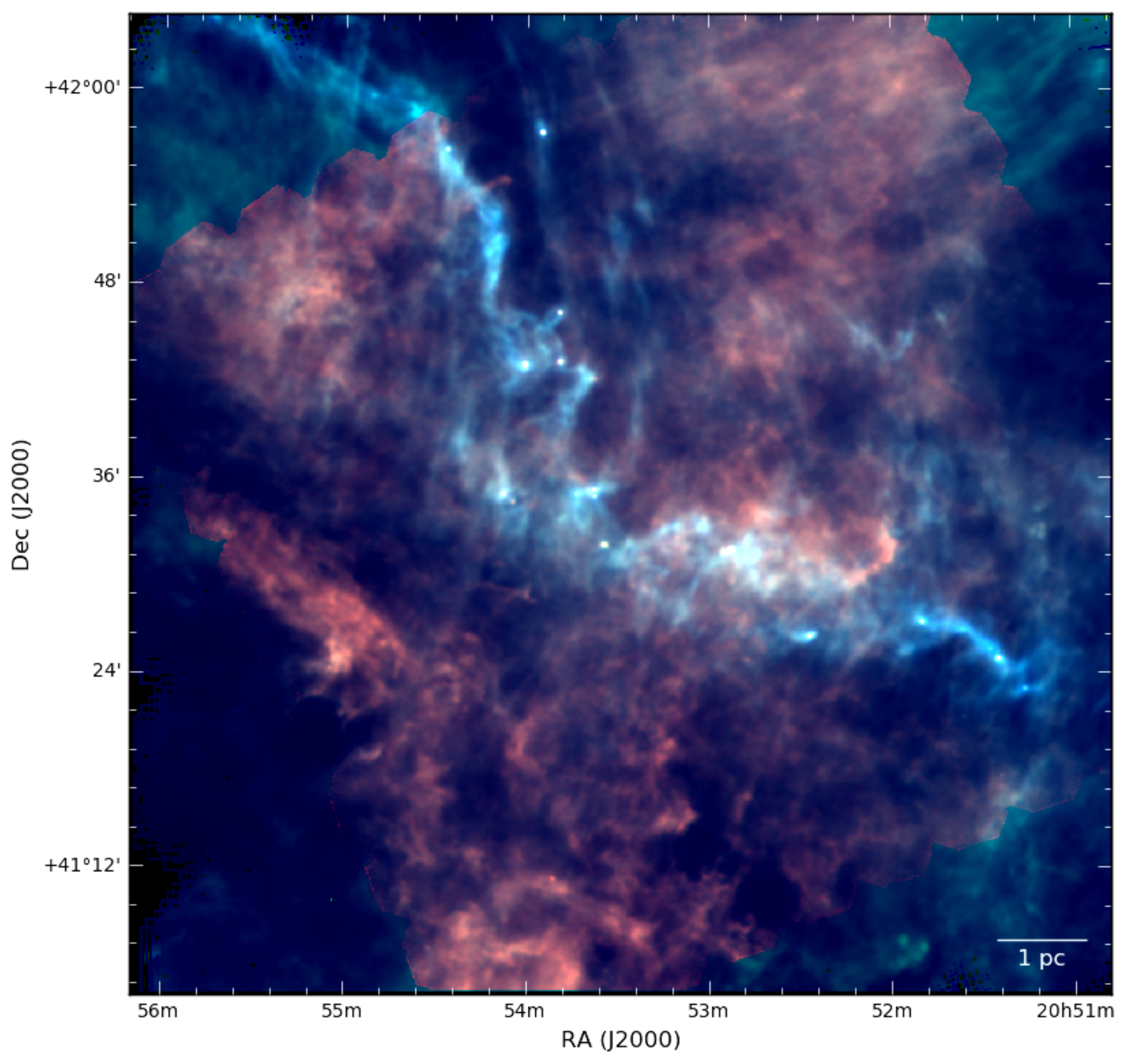}
\caption{An RGB image of the field G82.65-2.00. The colours correspond to $Herschel$ channels at 160 $\mu$m (red), 250 $\mu$m (green) and 350 $\mu$m (blue).}
\label{fig:RGB}
\end{figure*}

The $Herschel$ open time key programme Galactic Cold Cores (GCC) carried out dust continuum emission observations of 116 fields that were selected based on the $Planck$ Catalogue of Galactic Cold Clumps (PGCC). The fields were mapped with $Herschel$ PACS and SPIRE instruments \citep{Pilbratt2010,Poglitsch2010,Griffin2010} at wavelengths of $100-500 \mu$m. $Herschel$ makes it possible to study the $Planck$ clumps and filaments and their internal structure in detail \citep{Juvela2012,Montillaud2015,RiveraIngraham2015}. First results from $Planck$ and $Herschel$ have been presented in \citet{PlanckXXIII}, \citet{PlanckXXII}, and in \citet{Juvela2010,Juvela2011,Juvela2012}. \citet{Montillaud2015} present an analysis of submillimetre clumps and star formation in these $Herschel$ fields.

\begin{figure*}
\centering
\includegraphics[width=17cm]{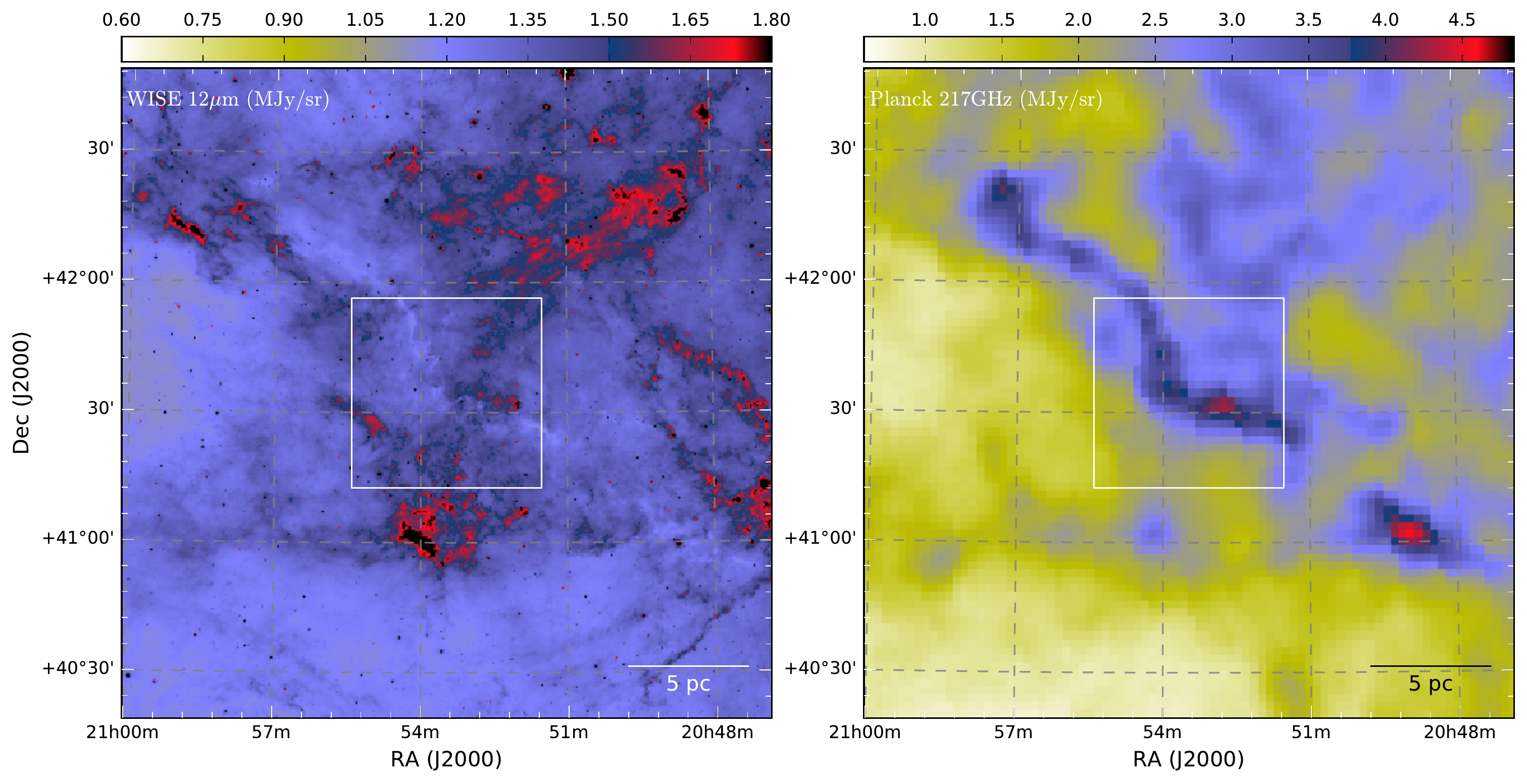} 
\caption{A $WISE$ 12 $\mu$m map (left) and a $Planck$ 217 GHz map (right). The $Planck$ map has been converted to MJy/sr using the conversion factors given in \citet{PlanckIX}. The white rectangle shows the extent of the maps in Fig. \ref{fig:coldmap}.}
\label{fig:warmmap}
\end{figure*}

\begin{figure*}
\centering
\includegraphics[width=17cm]{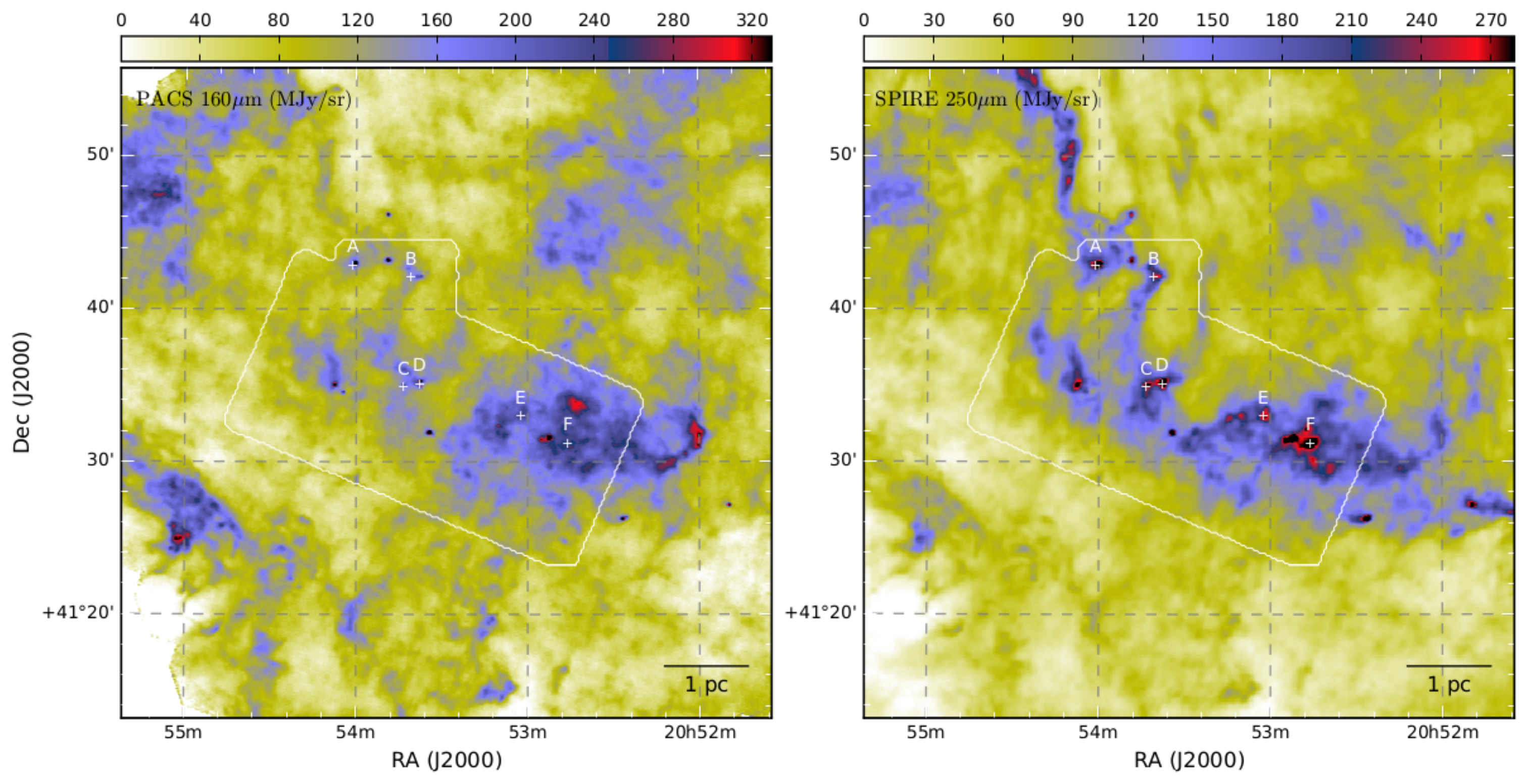} 
\caption{Surface brightness observed by $Herschel$: PACS 160 $\mu$m map (left), and SPIRE 250 $\mu$m map (right). The labels from A to F and the white crosses show the locations of cold clumps analysed in Section \ref{Sec3}. The white contour shows the area coverd by our line observations from the Nobeyama 45 m telescope.}
\label{fig:coldmap}
\end{figure*}

For this paper, we have selected one of the GCC fields, G82.65-2.00. The filament was identified from $Planck$ observations as a region of particularly cold dust emission, with the cold dust concentrated to a filamentary structure (Figs. \ref{fig:RGB}, \ref{fig:warmmap}), and \ref{fig:coldmap}. The observations of the field revealed a very prominent, strongly fragmented cloud filament, rich in young stellar objects (YSOs) and starless clumps. In total, 28 YSOs and 24 cores have been found to be associated with the main filament \citep{Montillaud2015}. With the assumed distance of 620 pc (see Section 3) the mass of the filament exceeds 1200 $M_{\odot}$. In this paper we utilize line observations of several molecules, including CO isotopologues, $\rm HCO^{+}$, HCN, and CS from the Osaka 1.85 m telescope and the Nobeyama 45 m telescope to add kinematical information to study possible correlations between the gas and dust components.

The morphology of G82.65-2.00 is filamentary, although fragmented. The strong fragmentation indicates that the filament is more evolved than, for example, the Taurus filaments B211/213 and TMC-1 or Musca. Compared to high mass star forming infrared dark clouds, it is less massive and isolated being located at $b = -2\degr$, suffering much less from confusion by other clouds on the same line-of-sight. The fragmented nature of the region, and the clear evidence of core formation, may infer that the substructures are dissolving back to the interstellar medium (ISM). If the filament is truly dissolving it would be a prime candidate to study the evolutionary stage of a 'debris filament', an evolutionary stage of filaments after star formation but before the structures have dissipated to the ISM.

The content of this paper is as follows. In Section 2, we give an overview of our observations. In section 3, we present our main results, and in Section 4, we provide some discussion concerning our main results. In Section 5, we give our concluding remarks.

\section{Observations}\label{Sec2}

\subsection{Dust continuum observations}

The $Herschel$ observations are discussed in detail by \citet{Juvela2012} and \citet{Montillaud2015}. The SPIRE observations at 250 $\mu$m, 350 $\mu$m, and 500 $\mu$m were reduced with the $Herschel$ Interactive Processing Environment HIPE v.12.0, using the official pipeline with the iterative destriper and the extended emission calibration options. The maps were produced with the naive map-making routine. The accuracy of the absolute calibration of the SPIRE observations is expected to be better than $7 \%$. The PACS data at 160 $\mu$m was processed with HIPE v.12.0 up to Level 1 and the final map was produced with Scanamorphos v.23 \citep{Roussel2013}. For PACS, when combined with SPIRE data, we adopt an error estimate of 10$\%$\footnote{\url{http://Herschel.esac.esa.int/twiki/bin/view/Public/PacsCalibrationWeb}}. The value is consistent with the differences between PACS and Spitzer MIPS measurements of extended emission. In order of increasing wavelength, the resolution is approximately 12$\arcsec$, 18$\arcsec$, 25$\arcsec$, and 37$\arcsec$ for the four bands. However, the total size of the filament is larger than our $Herschel$ maps. The densest ridge of the filament is visible in absorption, although faintly, even in the Wide-field Infrared Survey Explorer \citep[$WISE$,][]{Wright2010} 12 $\mu$m map in Fig. \ref{fig:warmmap}, extending over $\sim$25 pc in total.

\subsection{Large scale CO observations with the Osaka 1.85 m telescope}

In order to study the velocity field as well as the molecular distributions, we have observed an area covering $\sim$ $1.25^{\circ} \times 1.25^{\circ}$, corresponding to the white rectangle in Fig. \ref{fig:warmmap}, with the $\rm ^{12}CO$ ($J$=2-1), $\rm ^{13}CO$ ($J$=2-1), and $\rm C^{18}O$ ($J$=2-1) emission lines using the 1.85m Osaka telescope \citep{Onishi2013}. The observations were carried out in September 2013 in the same way as for the observations of Monkey Head nebulae reported by \citet{Shimoikura2013} and using the on-the-fly (OTF) mapping technique.

The half power beam width (HPBW) of the telescope is $2.7'$ at 230 GHz, the rest frequency of the $^{12}$CO($J=2-1$) emission line. We used a digital spectrometer providing the $\sim 54$ kHz frequency resolution which corresponds to $\sim 0.07$ km s$^{-1}$ velocity resolution at 230 GHz. The velocity range of about $\pm100$ km s$^{-1}$ was covered by the observations. Noise level of the resulting data is $\sim 0.6$ K in units of $T_{\rm mb}$. For more details of the telescope and the observational methods, see papers by \citet{Onishi2013} and \citet{Shimoikura2013}.

The C$^{18}$O data suffers from a poor signal-to-noise ratio (S/N), because the emission line is weak, and thus we will use only $^{12} \rm CO$ and $^{13} \rm CO$ data in our analyses. We display velocity channel maps of the $^{12} \rm CO$ and $^{13} \rm CO$ emission lines in Appendix A (Figs. \ref{fig:12CO_cont_osaka} and \ref{fig:13CO_cont_osaka}).

\subsection{Molecular line observations with the Nobeyama 45 m telescope}
Further spectral line observations were carried out with the 45 m telescope at the Nobeyama Radio Observatory (NRO) for 15 days in April and May 2014. We used the waveguide-type dual-polarization sideband-separating SIS receiver TZ \citep{Nakajima2013} with the digital spectrometer SAM45. The set-up provides 4096 channels with a $\sim$ 8 kHz frequency resolution covering an $\sim$ 32 MHz bandwidth. The corresponding velocity resolution and velocity coverage at 110 GHz are $\sim$ 0.025 km$\rm s^{-1}$ and $\sim$ 100 km$\rm s^{-1}$. The beam size of the telescope at 110 GHz is $\sim 14.2''$ (HPBW).

We observed the $^{12} \rm CO(\textit{J} = 1 - 0)$,  $^{13} \rm CO(\textit{J} = 1 - 0)$,  $\rm C^{18}O(\textit{J} = 1 - 0)$, $\rm CS(\textit{J} = 2 - 1)$, and $\rm HCO^{+}(\textit{J} = 1 - 0)$ emission lines. The combination of the TZ receiver and the SAM45 spectrometers, however, enables us to observe some additional molecular lines simultaneously (up to sixteen lines) with 8 to 16 GHz frequency separation. In total, we observed twenty molecular lines as summarized in Table \ref{tab:observations}.

We mapped an area of $12' \times 24'$ around the main filament using the OTF technique \citep{Sawada2008}, and calibrated the spectral data with the standard chopper-wheel method \citep{Kutner1981}. The reference position for the OTF observations (i.e., the emission-free OFF positions) was chosen to be ($ \alpha_{\rm J2000}, \delta_{\rm J2000}) = \rm (20h57m05.04s, \rm 39^{\circ}19'59.4'')$. During the observations, we discovered a well-defined small core located just outside of the $12' \times 24'$ area covering the main filament. Therefore, we carried out an additional OTF mapping covering a $\sim 2' \times 2'$ area around the small core in certain molecular lines (see Table \ref{tab:observations}). The extent of the Nobeyama observations is shown in Fig. \ref{fig:coldmap}.

We used the reduction software package NOSTAR available at NRO to subtract a linear baseline from the raw data and to re-sample the spectral data onto a $\rm 10''$ grid aligned with the equatorial coordinates. A varying main beam efficiency correction, in range of $\eta_{mb} = 29.2 - 46.6 \%$ depending on the observed frequencies, was applied to scale the data to units of $ T_{\rm mb}$.

The frequency channels were re-sampled onto a 0.025 km$\rm s^{-1}$ velocity grid, and then smoothed to $\rm 0.2 - 0.4 km \rm s^{-1}$ resolution to reduce the noise. Noise levels are in the range of $\Delta T_{\rm mb} = 0.1-0.5 \rm K$ for a velocity resolution of 0.2 km$\rm s^{-1}$, as summarized in Table 1. Among the twenty molecular lines we observed, nine of them were detected with S/N greater than 3. These lines are labelled 'Yes' in the last column of the table. We display velocity channel maps of the nine emission lines in a series of figures in Appendix B (Fig. B1 to Fig. B9).

The system noise temperature including the atmospheric attenuation varied between $140 - 250$ K depending on the observed frequency. Pointing accuracy was better than $\sim 10''$, and was checked by observing a $\rm SiO$ maser, T-Cep, at 43 GHz and a $\rm H_{2}O$ maser, W75N, at 22GHz every $1 - 2$ hours.

\begin{table*}
\caption{Observed molecular lines from Nobeyama}
\label{tab:observations}
\centering
\begin{tabular}{c c c c c}
\hline\hline
& & & & \\
Molecule & Transition & Rest frequency & $\Delta \rm T_{\rm mb}$\tablefootmark{\rm (1)} & Detection \\
& & (GHz) & (K) & \\
\hline
& & & & \\
$^{12}\rm CO$ & $ J = 1 - 0 $ & 115.271202 & 0.53 & Yes \\
$\rm CN$ & $ J = 1 - 0$; $ J = 3/2 - 1/2$; $ F = 5/2 - 3/2 $ & 113.490982 & 0.40 & No \\
$\rm CCS$ & $ N;$ $ J = 9; 8 - 8; 7$ & 113.410204 & 0.41 & No\\
$\rm C^{17}O$ &$ J = 1 - 0$ & 112.358988 & 0.35 & No \\
$\rm CH_{3}CN$ &$ J = 6(2) - 5(2)$; $F = 7 - 6$ & 110.375052 & 0.21 & No \\
$\rm ^{13}CO$&$ J = 1 - 0$ & 110.201353 & 0.20 & Yes\\
$\rm NH_{2}D$&$ J = 1(1; 1)0 - 1(0; 1)0+ $& 110.153599 & 0.22  & No\\
$\rm C^{18}O$&$ J = 1 - 0 $& 109.782173 & 0.19 & Yes\\
$\rm H_{2}CS$&$ J = 3(1; 3) - 2(1; 2)$ & 101.477885 & 0.21 & No\\
$\rm HC_{3}N$&$ J = 11 - 10$ & 100.076385 & 0.16 & Yes\\
$\rm SO$&$ N; J = 5; 4 - 4; 4$ & 100.029565 & 0.16 (0.15) & No\\
$\rm SO$&$ N; J = 2; 3 - 1; 2$ & 99.299905 & 0.21 (0.15) & Yes\\
$\rm CS$&$ J = 2 - 1$ & 97.980953 & 0.13 (0.13) & Yes\\
$\rm CH_{3}OH$&$ J = 2(0; 2) - 1(0; 1)A + +$ & 96.741377 & 0.16 & Yes\\
$\rm C^{34}S$&$ J = 2 - 1$ & 96.412961 & 0.15 & No\\
$\rm CH_{3}OH$&$ J = 2(1; 2) - 1(1; 1)A + +$ & 95.914310 & 0.14 & No\\
$\rm HCO^{+}$&$ J = 1 - 0$ & 89.188526 & 0.19 (0.13) & Yes\\
$\rm HCN$&$ J = 1 - 0$; $ F = 2 - 1$ & 88.6318473 & 0.18 (0.13) & Yes\\
$\rm SiO$&$ J = 2 - 1$; $v = 0 $& 86.846995 & 0.22 (0.16) & No\\
$\rm H^{13}CO^{+}$ &$ J = 1 - 0$ & 86.754288 & 0.18 (0.14) & No\\

\hline
& & & & \\
\end{tabular}
\tablefoot{ Values in the parentheses are those for the additional observations to map a small core, clump B in Fig. \ref{fig:location}, located beside the main filament. \\
\tablefoottext{$\rm 1$}{Typical one sigma noise level (rms) at the 0.2 km$\rm s^{-1}$ velocity resolution.}
}
\end{table*}

\section{Results}\label{Sec3}

\subsection{Distance estimate}

\begin{figure}
\centering
\resizebox{\hsize}{!}{\includegraphics[width=17cm]{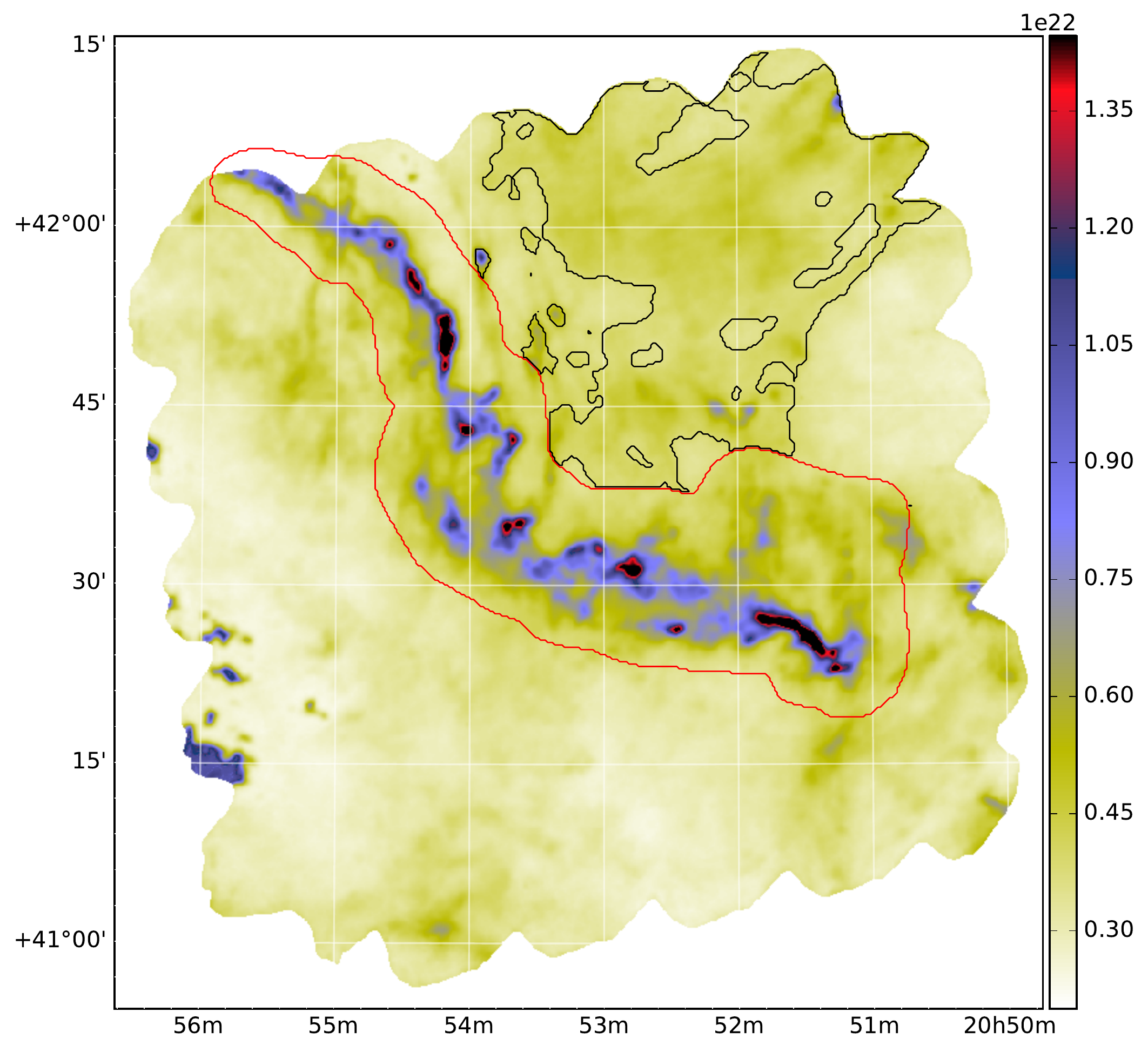}} 
\caption{The areas used by the extinction method to derive distance estimates. The area inside the red contour is used to estimate the distance of the main filament, and the area inside the black contour is used to estimate the distance of the secondary component.}
\label{fig:MACHETE}
\end{figure}

Distance estimate is crucial to derive the mass of molecular clouds and to investigate their kinematics. However, the distance to G82.65-2.00 has not been determined well up to date. \citet{Montillaud2015} estimated the distance to be $1 \pm 0.5$ kpc based on several methods but the scatter is large depending on the method. CO observations of \citet{Dame2001} show that the field around the filament consists of a single component at 3.47 km $\rm s^{-1}$, in agreement with our results showing the main line emission component in the range of $[3,5]$ km $\rm s^{-1}$ (see Section  \ref{ssec:3.2}). In general, distance can be derived from spectroscopic observations, assuming that the galactic gas is in circular rotation. Using the Galactic rotation curve by \citet{Reid2009, Reid2014}, \citet{Montillaud2015} derive two kinematics distances of $3.30 \pm 0.14$ and $1.08 \pm 0.76$ kpc assuming a tangential peculiar velocity of
 $V_0=0$ and $V_0=-15$ km $\rm s^{-1}$, respectively. This latter value corresponds to the average orbital velocity reported by \citet{Reid2009, Reid2014} for massive star forming regions. However it remains unclear whether $V_0=-15$ km $\rm s^{-1}$ is correct for G82.65-2.00.

Other studies of clouds that seem to be connected with G82.65-2.00 have reported similar distance estimates. \citet{Dobashi1994} proposed a distance of 800 pc for their cloud 16 ($l = 82.93^{\circ}$, $b = -2.03^{\circ}$), while \citet{Bally1980} adopted a distance of 1 kpc for the Pelican Nebula ($l = 84.63^{\circ}$, $b = +0.10^{\circ}$). A recent work by \citet{Cersosimo2007} shows that several structures are seen in the region of the Pelican nebula, corresponding to different velocity components. The closest cloud to G82.65-2.00 in their study ($l = 84^{\circ}$, $b = -1.7^{\circ}$), has similar velocity of 2.2 km $\rm s^{-1}$ for which \citet{Cersosimo2007} derived a distance of 0.5 - 0.7 kpc, and suggested greater distances (1.7 kpc, 2.7 kpc and 3.3 kpc) for the other structures.

Because of the large uncertainty in the kinematic distance estimate, \citet{Montillaud2015} derived an independent estimate using  extinction method. By comparing observed stellar colours from the Two Micron All Sky Survey \citep[2MASS]{2MASS} point source catalogue with the predictions of the Besançon Galaxy model \citep{Robin2003, Robin2012}, they derived the most probable three dimensional extinction distribution along the line-of-sight. The method is described by \citet{Marshall2006, Marshall2009}. \citet{Montillaud2015} applied the method to two circular areas with $5\arcmin$ and $10\arcmin$ in diameter, centred on the brightest position of the $Herschel$ 250 $\rm \mu m$ map, which corresponds to clump F in Fig. \ref{fig:coldmap}. The estimated distances were 0.42 kpc for the small aperture and 0.98 kpc for the large aperture. The difference between the two values is included in the formal uncertainties provided by the method.

\begin{figure*}
\centering
\includegraphics[width=17cm]{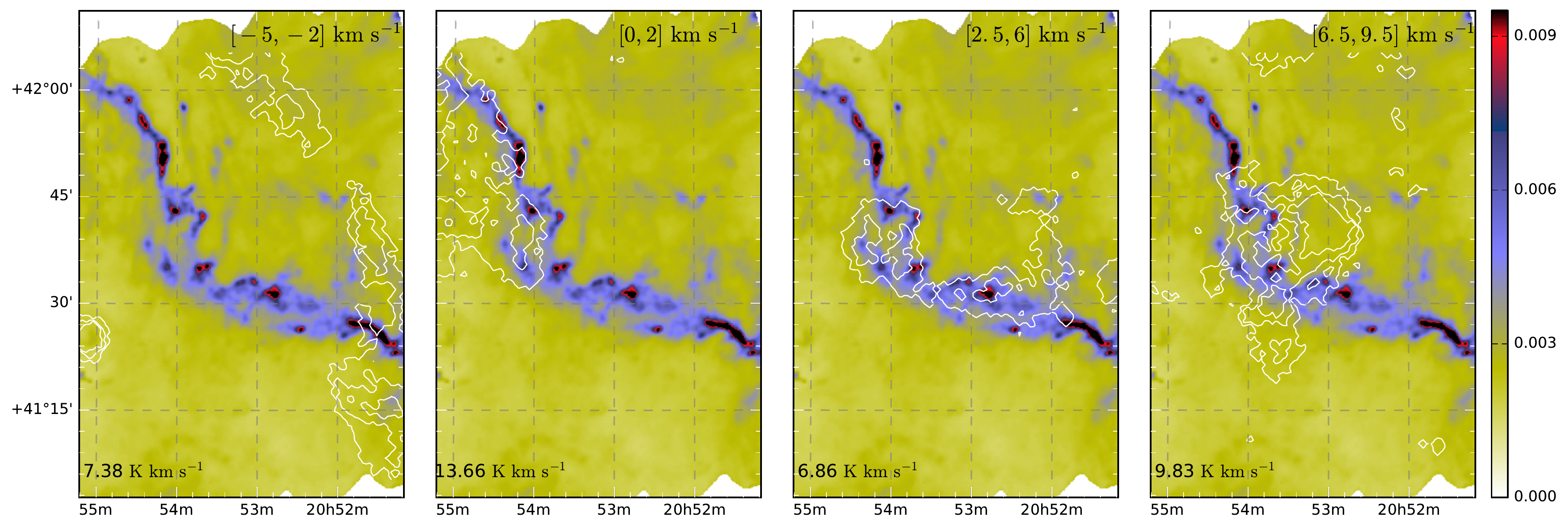} 
\caption{The different velocity structures identified from the Osaka observations. The underlying colour map is the $\tau_{250}$ map. The contours have been computed from the Osaka $^{12} \rm CO$ data and show 50$\%$, 70$\%$, and 90$\%$ of the peak emission in the velocity interval noted in the upper right corner. The peak emission in each velocity interval is denoted in the lower left corner.}
\label{fig:cartoon}
\end{figure*}

Considering the large uncertainties between the different methods, we decided to obtain a more reliable estimate for the distance using an updated version of the extinction method described above \citep[][Marshall et al. 2017 (in preparation)]{Marshall2015}. \citet{Montillaud2015} used only a narrow area around clump F, and thus the number of stars that can be used in the method was limited, which induced the large uncertainty in the distance estimate. We have increased the area in the extinction method to cover the entire cold filament as indicated by the red contour in Fig. \ref{fig:MACHETE}. Looking at the $Herschel$ data and the $WISE$ 12$\mu$m map (Figs. \ref{fig:RGB} and \ref{fig:warmmap}), it is likely that there is a second, warm and more extended filamentary structure that runs through the whole field from South-East to North-East. The structure is visible in Fig. \ref{fig:RGB} in red. We have used the extinction method to derive a distance estimate for this warmer filament using the area indicated by the black contour in Fig. \ref{fig:MACHETE}.

\begin{figure*}
\centering
\includegraphics[width=17cm]{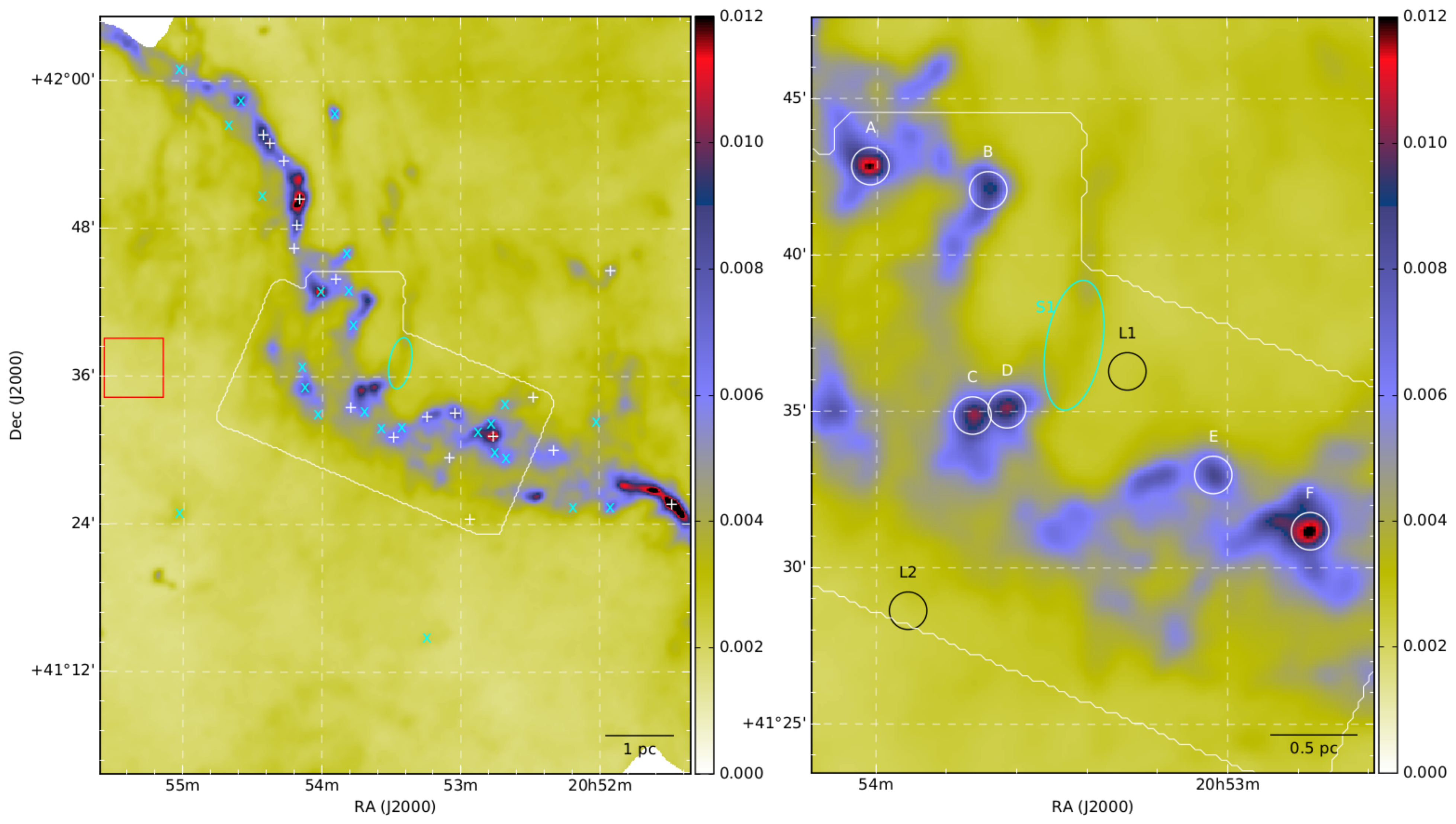} 
\caption{Left panel: the optical depth map of 250 $\mu$m derived from the $Herschel$ observations, the white plus signs indicate starless cores and the cyan crosses indicate protostellar source candidates identified by \citet{Montillaud2015}. Right panel: the clumps (labelled A to F) chosen for the spectral analysis (Figs. \ref{fig:spectra} and \ref{fig:spectra_rest}). The labels L1 and L2 indicate the positions where we have extracted spectra to analyse the warm filament (Fig. \ref{fig:third_12CO}). The white outline in the panels corresponds to the area of the Nobeyama observations. The cyan ellipse marks the selected striation S1, and the red rectangle marks the area used for the background subtraction as discussed in Section \ref{ssec:3.5}.}
\label{fig:location}
\end{figure*}

The computations were performed assuming either one or two clouds on the line-of-sight. For the cold filament, the highest posterior probability is obtained with one cloud at $620^{+31}_{-42}$ pc. For the warm filament, the results varied depending on the area to apply the method. With a smaller area, as seen in Fig. \ref{fig:MACHETE}, the highest probability is reached with one component at 300 pc. However, using a wider area the highest probability is obtained by two clouds, with one cloud at $661^{+69}_{-55}$ pc and a second cloud at $7.3^{+0.7}_{-1.6}$ kpc. A distance of 300 pc is plausible, however, a distance of $\sim$ 7 kpc is unlikely as the field is resolved with $Herschel$.

Thus, for the cold filament, we will adopt the distance 620 pc in this paper. The distance of the warm filament is more uncertain and requires further studies.


\subsection{Large-scale kinematics} \label{ssec:3.2}

The Osaka $\rm ^{12}CO$ data, Fig. \ref{fig:cartoon}, reveals an intriguing overall kinematic structure, with one filament visible in the range $\sim [-5,-2]$ km $\rm s^{-1}$ and extended emission appearing in the range $\sim [0,10]$ km $\rm s^{-1}$. At low radial velocities (see Fig. \ref{fig:12CO_cont_osaka} for more detailed velocity channel maps), in the range $\sim [0,4]$ km $\rm s^{-1}$, the eastern edge of the extended emission seems to trace well the cold filament. At velocities higher than 7 km $\rm s^{-1}$, there is a coherent velocity structure perpendicular to the central part of the cold filament. The large scale Osaka $\rm ^{12}CO$ data would thus imply that the cold filament seen in the $Herschel$ data is located between three, possibly four, different velocity structures.

The three compact sources seen in the lower left corner of the channel maps, Fig. \ref{fig:12CO_cont_osaka}, in the range $\sim [-6,1]$ $\rm km s^{-1}$ are probably related to the Infrared Astronomical Satellite (IRAS) point source 20529+4114. Although the fainter object in the range $\sim [-6,-4.5]$ $\rm km s^{-1}$ is within a few arc seconds of the IRAS source, the peak CO emission from the three sources is offset by $\sim 3 \arcmin$.

In the Osaka $\rm ^{13}CO$ maps (Fig. \ref{fig:13CO_cont_osaka}) the only prominent feature is a filament that matches the cold filament seen in the $Herschel$ data. The emission reaches maximum at $\sim 4$ km $\rm s^{-1}$ and concentrates around the main filament. At lower radial velocities, the emission is found in north-eastern and south-western part of the filament.

For the Nobeyama line observations, we show channel maps only in the velocity range $\sim [0.0,9.5]$ in Fig. \ref{fig:12CO_cont}. Although $^{12} \rm CO$ was also detected in the range of $\rm -4.7$ to $\rm -2.5$ km $\rm s^{-1}$, we have not included this velocity range in our analyses as it is distributed rather randomly, inferring there is no relation to the main filament. The channel maps of the Nobeyama observations are shown in Appendix B.

The Nobeyama $^{12} \rm CO$ emission in Fig. \ref{fig:12CO_cont} follows the same pattern as the Osaka data. The peak of the emission moves from the north-east towards south-west, tracing well the dust emission in each velocity interval. Some well defined substructures with different velocities can be identified, for example, in the range $\sim [-0.4,1.5]$ km $\rm s^{-1}$ and in the range $\sim [7,9]$ km $\rm s^{-1}$. 

As expected, and as seen in the data from Osaka, the $^{13} \rm CO$ line traces better the shape of the cold filament visible in dust emission in Fig. \ref{fig:coldmap}. Some individual substructures can be identified from the contour maps. In the range of $\sim [4,6]$ km $\rm s^{-1}$, the $^{13} \rm CO$ line clearly traces the striation S1, marked with a cyan ellipse in Fig. \ref{fig:location}, and connects the striation to the dense clumps C and D.

The Nobeyama $\rm C^{18}O$ data (Fig. \ref{fig:C18O_cont}) traces the densest region of the cold filament but the line is also seen on the striation S1 (see Fig. \ref{fig:location}) and, as with the Nobeyama $^{13} \rm CO$, the line emission extends to clumps C and D.

\begin{figure*}
\centering
\includegraphics[width=17cm]{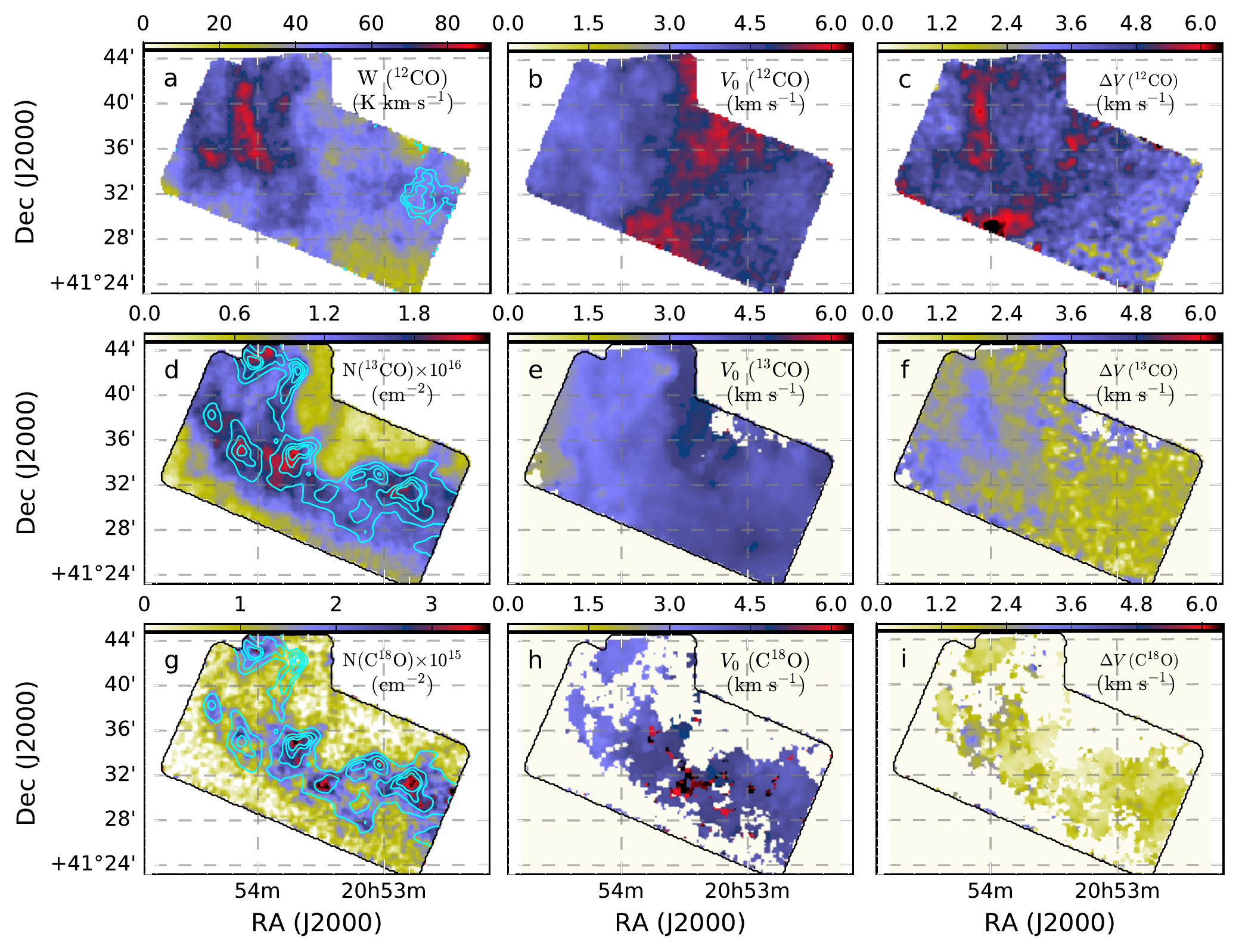} 
\caption{First row: The total integrated intensity (K km $\rm s^{-1}$) of $\rm ^{12}CO$, the velocity and the line width at FWHM (km $\rm s^{-1}$) obtained from the Gaussian fit of the $^{12}$CO line. Second and third rows: the total column densities and the best-fit values for velocity and the line width obtained from the Gaussian fit of the $\rm ^{13}CO$ and $\rm C^{18}O$ lines, respectively. The column densities are derived with a constant excitation temperature of 15K.  The contour in the $W$ $(\rm ^{12}CO)$ map shows the observed $\rm ^{12}CO$ emission in the range of $[-4,-1]$ km $\rm s^{-1}$ which was not included in our analysis. The lowest contour is at 0.7 K km $\rm s^{-1}$ and the contours increase with 0.7 K km $\rm s^{-1}$ steps. The contours in the $\rm ^{13}CO$ and $\rm C^{18}O$ column density maps correspond to 45$\%$, 60$\%$, 75$\%$, and 90$\%$ of the maximum value of $\tau_{250}$. We have masked regions with S/N < 3 in the velocity and line width maps of $\rm ^{13}CO$ and $\rm C^{18}O$.}
\label{fig:map_plot}
\end{figure*}

In order to study the gas dynamics, we computed pixel-by-pixel Gaussian fits for the Nobeyama $\rm ^{12}CO, ^{13}CO$, and $\rm C^{18}O$ data. To reduce the noise level in the resulting images, the data were averaged over $30'' \times 30''$ corresponding to $3 \times 3$ map pixels. The Gaussian fits returns the three line parameters: the peak brightness temperature $T_{\rm mb}$, the radial velocity $V_{\rm 0}$, and the the line width $\Delta V$ measured at FWHM. The results for $V_{\rm 0}$ and $\Delta V$ are shown in the second and third columns in Fig. \ref{fig:map_plot}. Because of the different velocity components, the shape of the $\rm ^{12}CO$ line is complex, and thus, a single Gaussian fit may not describe the properties of the line well. However, results described below do not change even when we fit the spectra with two Gaussian components.

Figure \ref{fig:map_plot}a shows the observed total integrated intensity of $\rm ^{12}CO$. The emission reaches maximum on the Eastern side of the image, and traces the same region as the velocity component seen in the range $\sim [-0.5,2]$ km $\rm s^{-1}$ in Fig. \ref{fig:12CO_cont}. The overlaid contours show the distinct emission component in the range $\sim [-4,-1]$ km $\rm s^{-1}$ that is not used in the analysis. The velocity map in Fig. \ref{fig:map_plot}b shows that the peak velocities are in the range $\sim [3,5]$ km $\rm s^{-1}$, with a notable exception in the central part of the panel. The higher velocity component corresponds to the velocity structure seen in the range $\sim [7,9]$ km $\rm s^{-1}$ in Fig. \ref{fig:12CO_cont}. The two velocity components are also clearly distinguishable in the line width map in Fig. \ref{fig:map_plot}c, but the larger line width also corresponds to the eastern part of the filament, where the integrated emission peaks.

The velocity and line width maps of $\rm ^{13}CO$ and $\rm C^{18}O$ (Fig. \ref{fig:map_plot}f and \ref{fig:map_plot}i) show the main emission in the range $\sim [3,5]$ km $\rm s^{-1}$. However the features seen in the $\rm ^{12}CO$ maps, the clearly distinguishable higher velocity component and two components in the line width map, are not seen, because $\rm ^{13}CO$ and $\rm C^{18}O$ are not detected at velocities higher than $\sim 5.5$ km $\rm s^{-1}$. The white areas in the maps were masked out because of low signal-to-noise ratio. 

\subsection{Column density estimates from line observations and dust emission}

\begin{figure}
\centering
\resizebox{\hsize}{!}{\includegraphics[width=17cm]{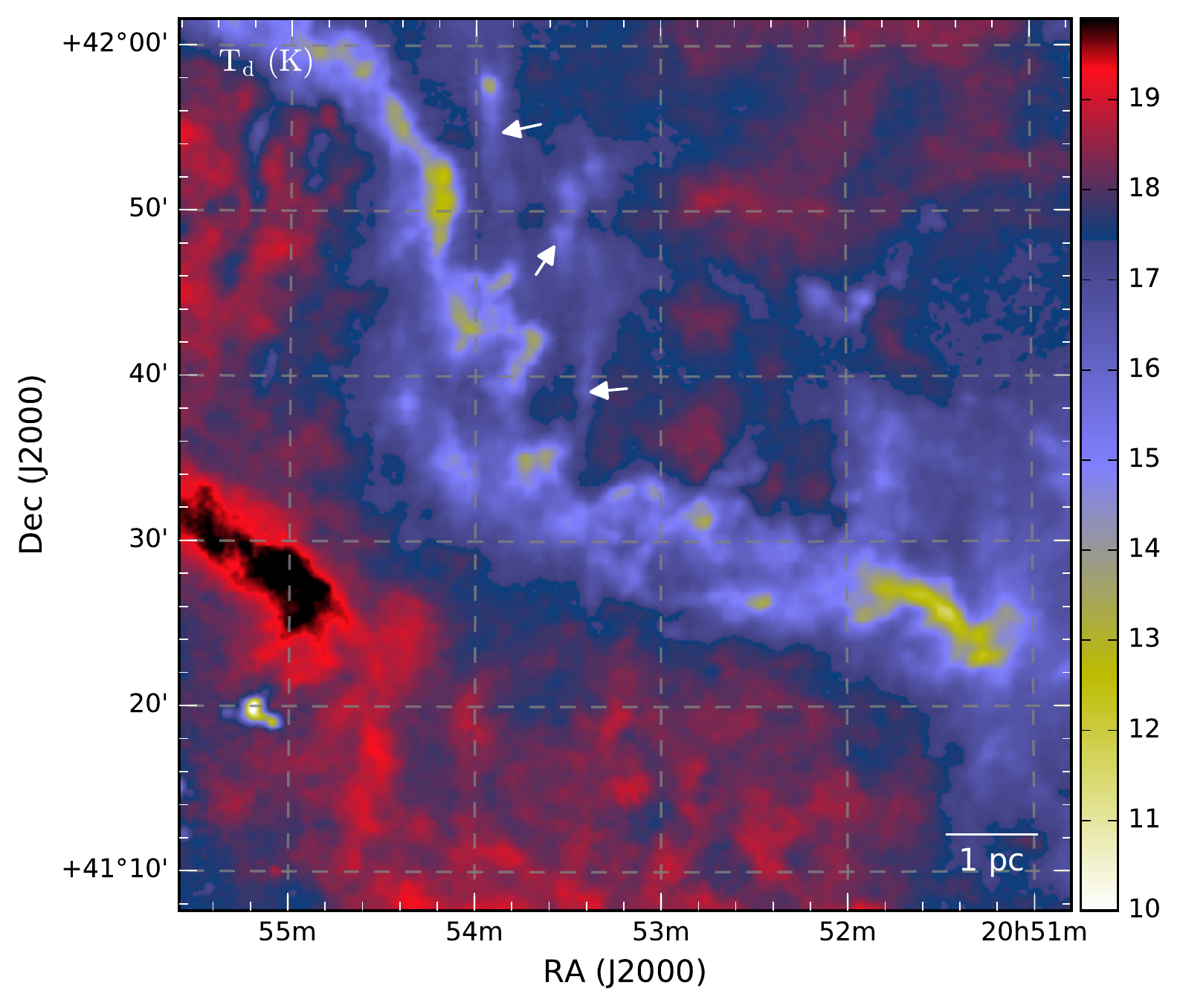}} 
\caption{A dust temperature map derived from the $Herschel$ observations. The map is truncated at 19.9 K, to show the fainter structures. The white arrows indicate structures identified as striations.}
\label{fig:Tmap}
\end{figure}

Shown in Fig. \ref{fig:Tmap}, is a dust temperature map derived from the $Herschel$ observations. The temperature was estimated by fitting the spectral energy distributions (SEDs) with modified blackbody curves with a constant opacity spectral index of $\beta = 1.8$, which is the average value in molecular clouds \citep{PlanckXXII, PlanckXXIII, PlanckXXV, Juvela2015VI, Juvela2015V}. The cold filament, seen in blue in Fig. \ref{fig:Tmap}, contains substructures of a few times 0.1 pc including several cold clumps. In the $Herschel$ images, one can identify a number of fainter striations some of which are indicated by white arrows in Fig. \ref{fig:Tmap}.

We use the SPIRE observations at 250 $\mu$m, 350 $\mu$m, and 500 $\mu$m, to estimate the dust optical depth at 250 $\mu$m. The surface brightness maps were convolved to a common resolution of 40$\arcsec$, and colour temperatures were calculated by fitting the spectral energy distributions (SEDs) with modified blackbody curves with a constant opacity spectral index of $\beta = 1.8$. The 250 $\mu$m optical depth was obtained from 

\begin{equation}
\tau_{\rm 250} = \frac{I_{\nu}(\rm 250 \mu m)}{B_{\nu}(T_{\rm c})},
\end{equation}
using the fitted 250 $\mu$m intensity $I_{\nu}(\rm 250 \mu m)$ and the colour temperature $T_{\rm c}$. The data were weighted according to $7 \%$ error estimates for SPIRE surface brightness measurements.

The $Herschel$ observations are used to derive an estimate of the total hydrogen column density. The optical depth $\tau_{\nu}$ at frequency $\nu$ can be expressed as

\begin{equation}
\tau_{\nu} = \kappa_{\nu} N(\rm H_2) \hat{\mu},
\end{equation}
where $\hat{\mu}$ is the total gas mass per $\rm H_2$ molecule, 2.8 amu, and $N(\rm H_2)$ is the hydrogen column density. We assume a dust opacity of $\kappa_{\nu} = 0.1(\nu / 1000$ $\rm GHz)^{\beta}$ $ \rm cm^2 g^{-1}$ \citep{Beckwith1990}. The resulting optical depth map is shown in Fig. \ref{fig:location}.

The Nobeyama $\rm C^{18}O$ data is used to derive an independent estimate for the hydrogen column density. The column density $N$ of $\rm C^{18}O$ is derived following \citet{Garden1991} as

\begin{multline}\label{eq:column_density}
N = \frac{3k_{b}}{8 \pi^3 \rm{B} \mu^2_{\textit{r}}} 
\frac{\rm{exp}(\textit{h}\rm{B} \textit{J(J+1)}/\textit{k}_{\textit{b}} \textit{T}_{ex})}{J+1} \\ \times \frac{\rm \textit{T}_{ex} +\textit{h}B/3\textit{k}_{\textit{b}}}{\rm 1 - exp(-\textit{h}\nu / \textit{k}_{\textit{b}} \textit{T}_{\rm ex})}\int\tau_{\rm V} dV,
\end{multline} 
where B and $\mu_r$ are the rotational constant and the permanent dipole moment of the molecule, $J$ is the rotational quantum number of the lower energy level ($J=0$), $h$ is the Planck constant, $k_b$ is the Boltzmann constant, and $T_{\rm ex}$ is the excitation temperature.

For the integral of the optical depth, we have followed \citet{Buckle2010}, with the exception that we do not use an averaged optical depth over the whole region, but rather compute the optical depth $\tau$ separately for each pixel and sum over the velocity range where we have detected $ \rm C^{18}O$,

\begin{equation}
\int \tau_{\rm V} dV = \frac{1}{J(T_{ex}) - J(T_{bg})} \frac{\tau}{1-e^{-\tau}} \int T_{b} dV,
\end{equation} 
where $T_{bg}$ is the cosmic background temperature 2.73 K, $T_b$ is the observed brightness temperature of a given pixel, and $J(T)$ is the temperature equivalent to the temperature $T$ in the Rayleigh-Jeans law \citep[e.g.][]{Garden1991}. The excitation temperature estimated from the peak brightness temperature of the Nobeyama $^{12}$CO data varies over the observed region (up to $\sim40$ K) with an average value of $\sim15$ K. In this paper, we assume a flat excitation temperature of $T_{ex}=15$ K. Note that the estimates of the molecular column densities are not very sensitive to $T_{ex}$, and assuming $T_{ex}=40$ K produces a higher column density by only a factor of $\sim 2$.

We use the column density $N({\rm C^{18}O})$ to derive an estimate for the total hydrogen column density $N(\rm H_2)$. We assume a relative abundance of $1.0 \times 10^{4}$ between $\rm H_2$ and $^{12} \rm CO$ and a relative abundance of $4.9 \times 10^2$ between $^{16} \rm O$ and $^{18} \rm O$ \citep{Garden1991}. Thus, the resulting hydrogen column density is

\begin{equation}
N(\rm H_2) = 4.9 \times 10^6 \textit{N}(\rm C^{18}O).
\end{equation} 
For comparison, using the column density $N({\rm ^{13}CO})$ results in hydrogen column densities within $\pm 10 \%$ of the values derived using $N({\rm C^{18}O})$.

The column densities of both $\rm ^{13}CO$ and $\rm C^{18}O$ (Fig. \ref{fig:map_plot}d and \ref{fig:map_plot}g) trace well the maxima of the $\tau_{250}$ distribution. The total hydrogen column density derived from the line observations is $1.0 - 2.3 \times 10^{22}$ $\rm cm^{-2}$, in very good agreement with the values derived from the $Herschel$ observations, $0.5 - 2.5 \times 10^{22}$ $\rm cm^{-2}$.

\subsubsection{Correlation between gas and dust}

\begin{figure*}
\centering
\includegraphics[width=17cm]{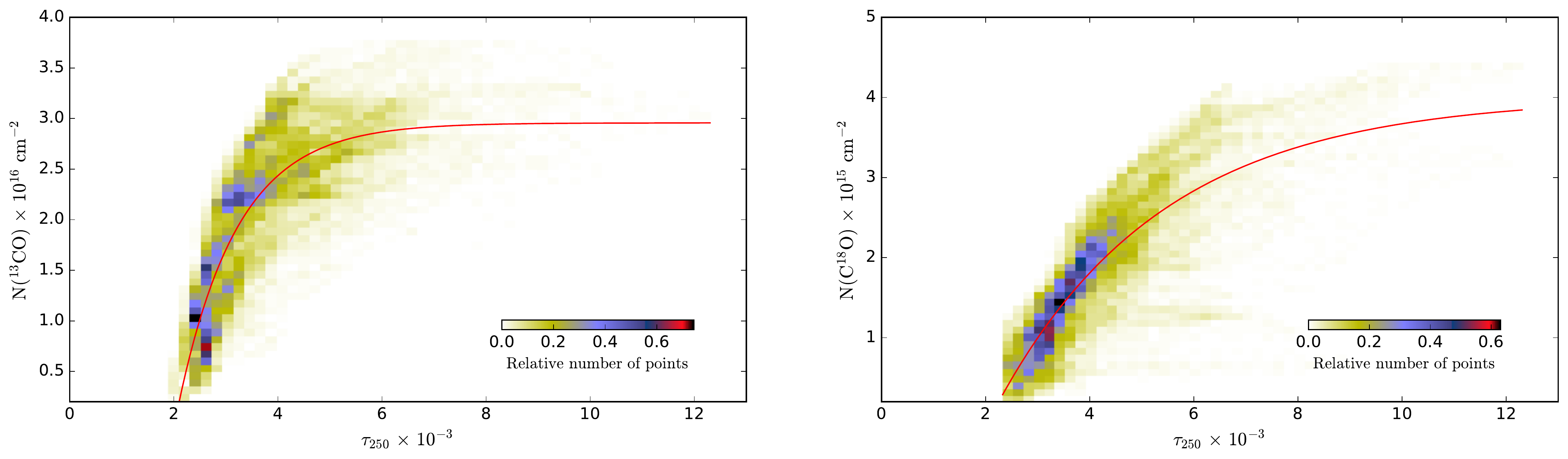} 
\caption{The pixel-to-pixel correlation of $\tau_{250}$ and the column density of $\rm ^{13}CO$ and $\rm C^{18}O$ derived from the Nobeyama observations. The red curve is the best fit of Eq. (\ref{eq:comp_fit}). The colour map corresponds to the density of data points.}
\label{fig:comp_2D}
\end{figure*}

Figure \ref{fig:comp_2D} shows the two-dimensional histograms of the pixel-to-pixel correlation between $\tau_{250}$ and the column densities of $\rm ^{13}CO$ and $\rm C^{18}O$ derived from the Nobeyama observations. The column density maps from the molecular line observations have been smoothed to the same resolution, 40$\arcsec$, as the $\tau_{250}$ map.  It is evident that the relation between $\tau_{250}$ and $N( \rm ^{13}CO)$ is not linear. We use a fit of the form,

\begin{equation}\label{eq:comp_fit}
N = N_0 (1 - \exp[- \rm c(\tau - \tau_{\rm CO})],
\end{equation}
where $N_0$ is the column density of the CO species at saturation, $\tau_{\rm CO}$ is the minimum dust optical depth for CO emission, and $c$ is a scaling factor. The results of the fits are shown in Fig. \ref{fig:comp_2D} and in Table \ref{tab:comp_fit_params}.

\begin{table}
\caption{Parameters of the fits to the $^{13} \rm CO$ and $\rm C^{18}O$ column density as a function of $\tau_{250}$.}
\label{tab:comp_fit_params}
\centering
\begin{tabular}{c c c c}
\hline\hline
& & & \\
Molecule & $N_0$ & $\tau_{\rm CO}$ & $c$ \\
\hline
& & & \\

$^{13}\rm CO$ & 2.96$\pm$0.07 & 0.88$\pm$0.09 & 2.02$\pm$0.8 \\
$\rm C^{18}O$ & 4.01$\pm$0.3 & 0.31$\pm$0.05 & 2.09$\pm$0.1 \\

\hline
& & & \\
\end{tabular}
\tablefoot{Best-fit values from Eq. (\ref{eq:comp_fit}) to data presented in Fig. \ref{fig:comp_2D}. }
\end{table}

The $^{13} \rm CO$ column density starts to rise rapidly after the optical depth reaches a threshold value of $ 2 \times 10^{-3}$, and starts to saturate when the optical depth reaches a value of $5 \times 10^{-3}$. On the other hand, the $\rm C^{18}O$, emission starts to rise rapidly at a slightly higher threshold value of $\tau_{250}$ $\approx$ $2.5 \times 10^{-3}$. Also, the correlation between $N (\rm C^{18}O)$ and $\tau_{250}$ is more linear below $\tau_{250}$ $\approx$ $5.5 \times 10^{-3}$. Furthermore, the column density of $\rm C^{18}O$ does not reach complete saturation.

\subsubsection{Position-velocity diagrams}

\begin{figure}
\centering
\resizebox{\hsize}{!}{\includegraphics[width=17cm]{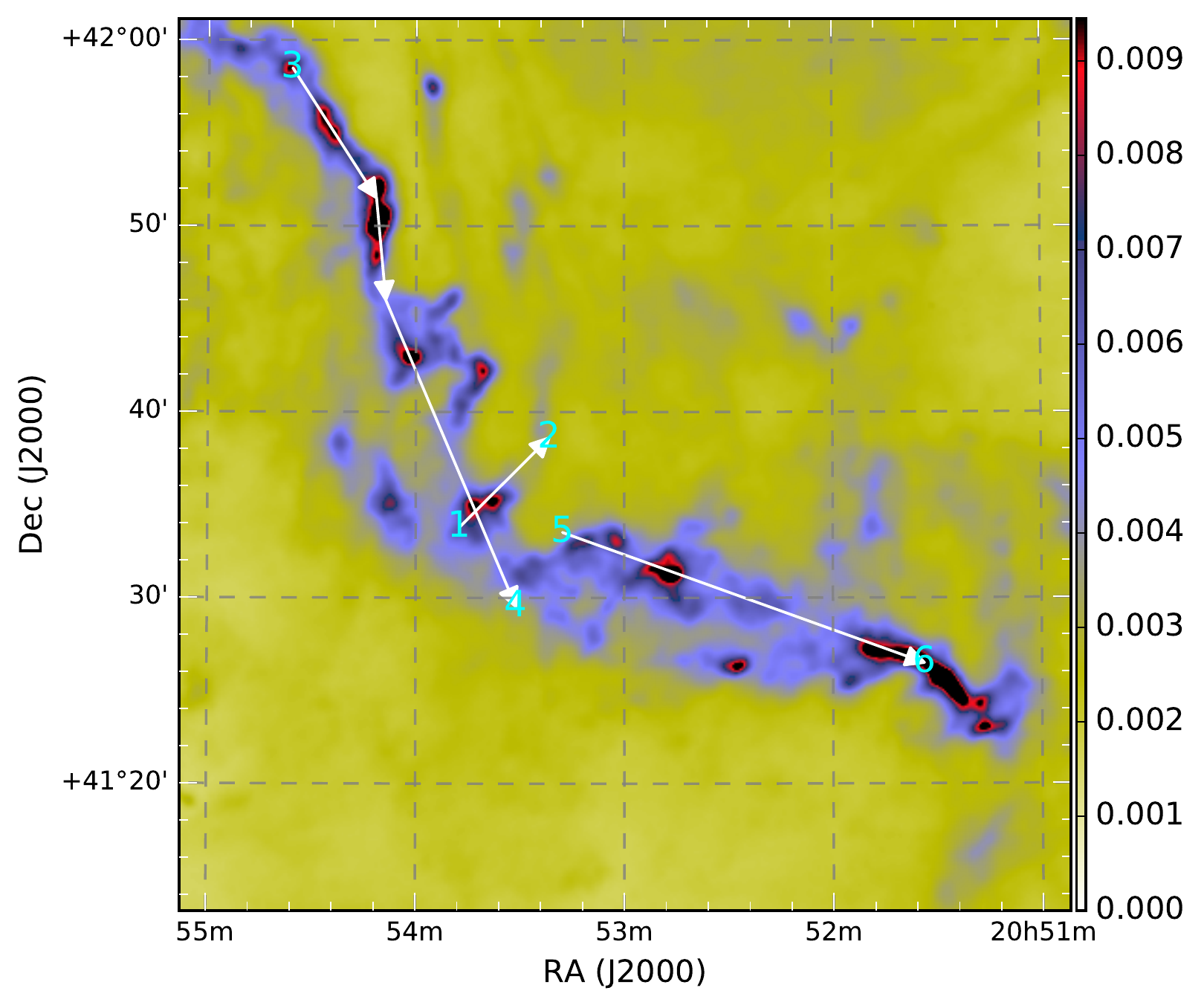}} 
\caption{The cuts used for the position-velocity maps, shown on the $\tau_{250}$ map.}
\label{fig:PV}
\end{figure}

\begin{figure}
\centering
\resizebox{\hsize}{!}{\includegraphics[width=17cm]{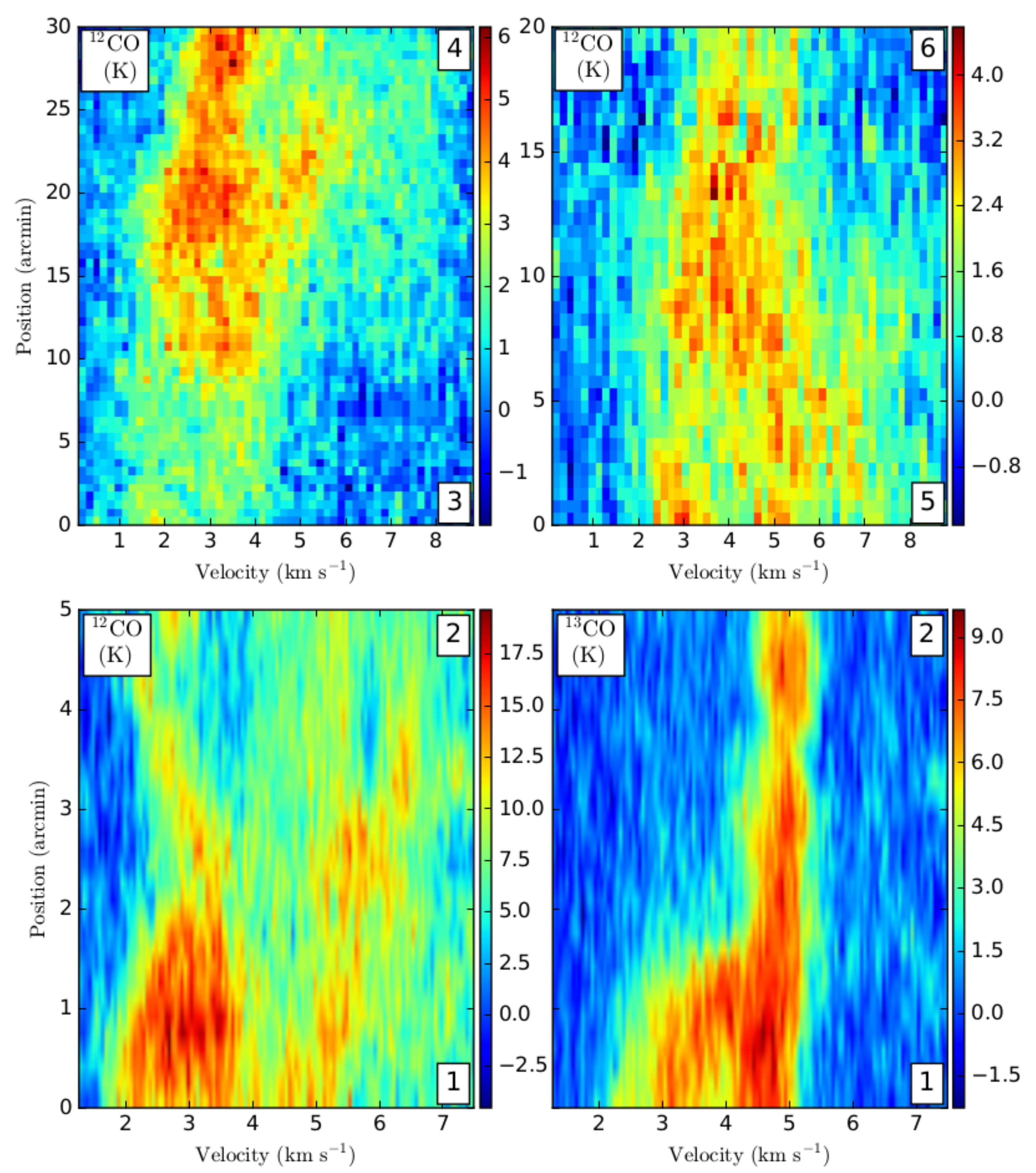}} 
\caption{Position-velocity maps, derived using the Osaka $^{12}$CO ($J=2-1$) data, upper row, and the Nobeyama $^{12}$CO ($J=1-0$) and $^{13}$CO ($J=1-0$) data, lower row. The numbers in the panels refer to the positions shown in Fig. \ref{fig:PV}.}
\label{fig:PV_maps}
\end{figure}

To study velocity gradients along the cold filament and possible signs of mass accretion, we have drawn position-velocity (PV) diagrams along the three cuts shown in Fig. \ref{fig:PV}. We have used the Osaka $^{12} \rm CO$ data to make two PV diagrams, from position 3 to 4 and from position 5 to 6. The diagrams are shown in Fig. \ref{fig:PV_maps}. The two segments of the cold filament are uniform, although, there are some substructures close to positions 4 and 5. However, due to low S/N it is not clear whether the velocity components are separate components, possibly at different distance on the same line-of-sight, or if the velocity components are a result of hierarchical fragmentation as discussed by \citet{Hacar2013_Bundles}. 

The Nobeyama $^{13} \rm CO$ and $\rm C^{18}O$ data indicate that the striation S1 is connected to the main filament. A PV diagram drawn along the striation from position 1 to position 2, Fig. \ref{fig:PV_maps}, shows a single velocity component between 4.0 and 5.5 km $\rm s^{-1}$. Thus, the striation is very likely connected kinematically to clumps C and D (see Fig. \ref{fig:location}).

\subsection{Mass estimates for clumps and substructures} \label{ssec:3.5}

\begin{figure}
\centering
\resizebox{\hsize}{!}{\includegraphics[width=17cm]{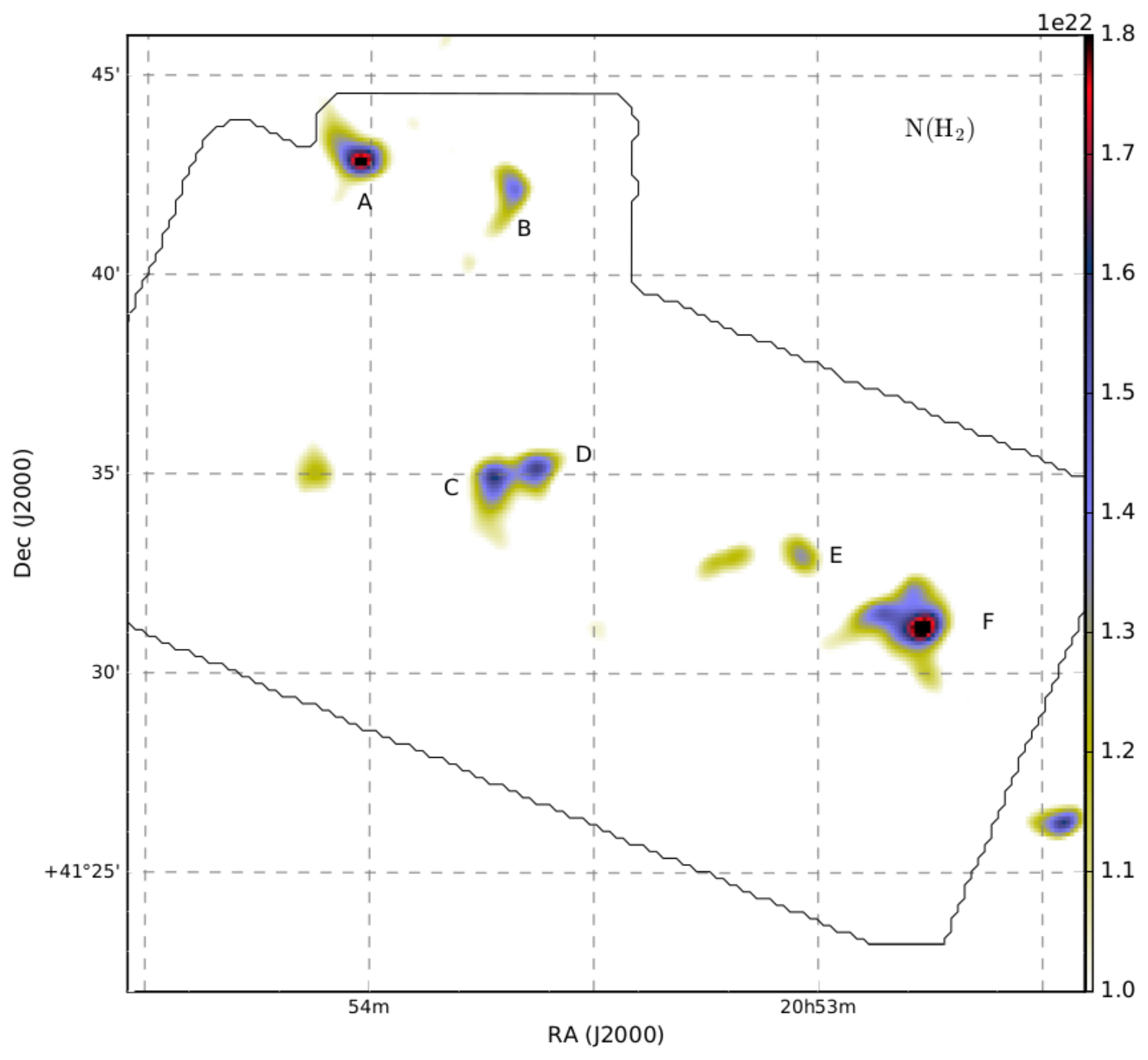}} 
\caption{A masked $\rm H_2$ column density map, showing the locations and column density of the clumps (labelled A to F) for which we have computed a mass estimate. The black contour shows the area of the Nobeyama observations.}
\label{fig:clumps}
\end{figure}

We have identified clumps as dense regions inside the area covered by the Nobeyama observations where the total hydrogen column density derived from dust is greater than $1.5 \times 10^{22}$ $\rm cm^{-2}$. In total, we have identified six clumps as shown in Fig. \ref{fig:clumps} and in the right panel of Fig. \ref{fig:location}. The mass of a clump can be written as

\begin{equation}\label{eq:mass_clump}
M_{\rm c} = N(\rm{H_2}) \pi \textit{R}^2  \hat{\mu} ,
\end{equation}
where $R$ is the radius of the clump, defined as half of the full-width at half maximum (FWHM) measured from the $Herschel$ column density map. The average molecular density $n$ of a clump is calculated using the average hydrogen column density $\overline{N}$ inside the radius as

\begin{equation}
n = \frac{\overline{N} (\rm H_2)}{2 \textit{R}}.
\end{equation}

To estimate the virial mass, we follow \citet{MacLaren}

\begin{equation}
M_{\rm virial} = \frac{k \sigma^2 R}{\rm G},
\end{equation}
where $k$ depends on the assumed density distribution, and G is the gravitational constant. We have chosen a density distribution corresponding to $\rho(\rm r) \propto \rm r^{-1.5}$, and thus, $k \approx 1.33$. The velocity dispersion, $\sigma$, can be computed from the equation

\begin{equation}
\sigma = \sqrt{\frac{k_b T_{\rm kin}}{\mu} + \frac{\Delta V^2}{8 \ln 2} - {\frac{k_b T_{\rm kin}}{m}}},
\end{equation}
where $\Delta V$ is the line width (at WFHM) of the line used, $m$ is the mass of the molecule used in the computation ($\rm ^{13} CO$), and $\mu$ is the mean molecular mass, 2.33 amu assuming $10\%$ helium. We have assumed a constant kinetic temperature of $T_{\rm kin} = 15$ K. The velocity dispersion can be used to estimate the nonthermal motion that provides support against gravity. If the mass of the cloud is less than the virial mass, it is not gravitationally bound and, without external pressure, will disperse.

The average hydrogen column density, average density, velocity dispersions for $\rm ^{13}CO$ and $\rm C^{18}O$, estimated mass, and virial mass estimates for the clumps are listed in Table \ref{tab:clump_masses}. The densities and masses are derived from the continuum observations. For the computations, a distance of 620 pc was assumed. The error estimates were calculated with Monte Carlo, assuming normally distributed errors.

To avoid possible contribution by the background in our analysis, a constant background value of $1.2 \times 10^{21}$ $\rm cm^{-2}$ was subtracted from the column density values when analysing dust continuum. The background value was estimated from the $Herschel$ images as the mean value inside the red rectangle in Fig. \ref{fig:location}. Thus, we do not take into account any possible contribution from the medium surrounding the clump. However, even for the more extended clumps, mainly clump F, the contribution from surrounding medium is at most $\sim 15\%$.

\begin{table*}
\caption{The average hydrogen column density, average density, velocity dispersion, estimated clump mass, and virial mass for the clumps shown in Fig. \ref{fig:clumps}. A distance of 620 pc is assumed.}
\label{tab:clump_masses}
\centering
\begin{tabular}{c c c c c c c c}
\hline\hline
& & & & & & &\\
 & $N(\rm H_2)$\tablefootmark{\rm (1)} & $n(\rm H_2)$\tablefootmark{\rm (1)} & $\sigma_{\rm ^{13} CO}$ & $\sigma_{\rm C^{18}O}$ & $M_c$\tablefootmark{\rm (1)} & $M_{\rm virial}(\rm ^{13}CO)$ & $M_{\rm virial}(\rm C^{18}O)$       \\
Clump & $ (10^{22}$ $\rm cm^{-2})$ & $( 10^4$ $\rm cm^{-3})$ & (km $\rm s^{-1}$) & (km $\rm s^{-1}$) & $(M_{\odot})$ & $(M_{\odot})$ & $(M_{\odot})$ \\
\hline
& & & & & & &\\

A & 1.54 $\pm$ 0.04 & 1.93 $\pm$ 0.05 & 0.68 $\pm$ 0.02 & 0.53 $\pm$ 0.01 & 15.2 $\pm$ 0.41 & 18.5 $\pm$ 0.94 & 11.4 $\pm$ 0.94 \\
B & 1.31 $\pm$ 0.03 & 2.07 $\pm$ 0.06 & 0.61 $\pm$ 0.03 & 0.52 $\pm$ 0.02 &  8.1 $\pm$ 0.22 & 11.7 $\pm$ 0.62 &  8.6 $\pm$ 0.94 \\
C & 1.46 $\pm$ 0.05 & 2.11 $\pm$ 0.05 & 0.88 $\pm$ 0.03 & 0.61 $\pm$ 0.03 & 10.6 $\pm$ 0.38 & 27.0 $\pm$ 0.84 & 15.6 $\pm$ 0.94 \\
D & 1.44 $\pm$ 0.04 & 2.26 $\pm$ 0.07 & 0.58 $\pm$ 0.02 & 0.41 $\pm$ 0.01 &  8.8 $\pm$ 0.36 & 10.8 $\pm$ 0.71 &  5.4 $\pm$ 0.94 \\
E & 1.25 $\pm$ 0.02 & 2.68 $\pm$ 0.08 & 0.61 $\pm$ 0.01 & 0.42 $\pm$ 0.02 &  4.2 $\pm$ 0.17 &  8.7 $\pm$ 0.53 &  4.3 $\pm$ 0.94 \\
F & 1.65 $\pm$ 0.04 & 1.93 $\pm$ 0.04 & 0.62 $\pm$ 0.02 & 0.51 $\pm$ 0.02 & 18.5 $\pm$ 0.52 & 16.4 $\pm$ 0.83 & 11.3 $\pm$ 0.94 \\

\hline
& & & & & & &\\
\end{tabular}
\tablefoot{(1) Estimated from dust emission.}
\end{table*}

\begin{figure}
\centering
\resizebox{\hsize}{!}{\includegraphics[width=17cm]{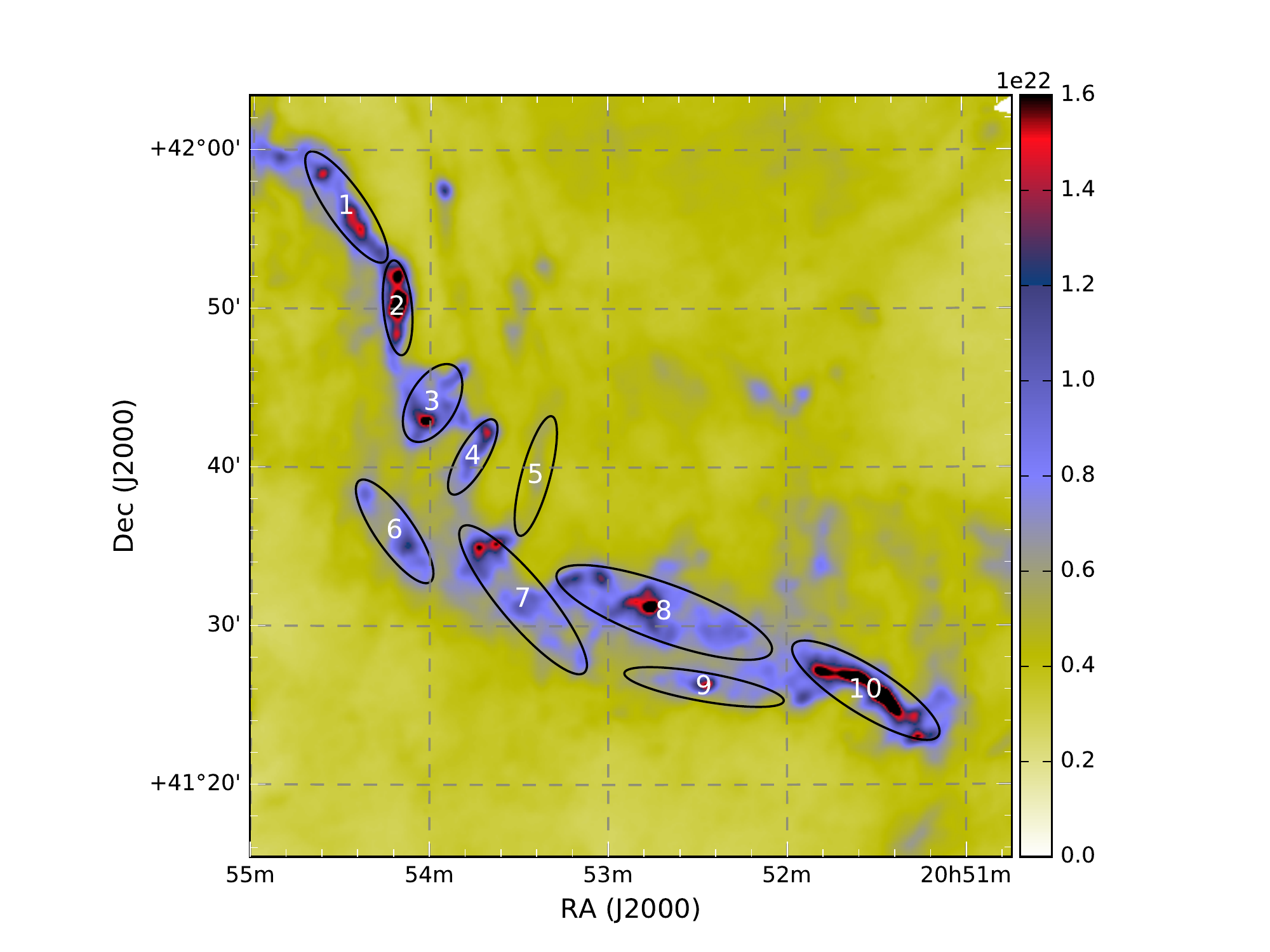}} 
\caption{The locations and surface areas of the identified substructures (labelled 1 to 10). The underlying colour map is the column density of $\rm H_2$ derived from the dust observations.}
\label{fig:filaments}
\end{figure}

We have identified some substructures within the filaments by eye, as labelled in Fig. \ref{fig:filaments}. These were first identified from the Osaka $\rm ^{13} CO$ data and compared with the $Herschel$ column density maps to identify the structures that are well correlated between gas and dust observations.

The substructure masses were estimated following Eq. (\ref{eq:mass_clump}), with the exception of radius which is defined as $R = \sqrt{S/\pi}$, where $S$ is the surface area of the structure shown by the ellipses in Fig. \ref{fig:filaments}. The surface area is computed as $S = \pi R_{\rm maj}R_{\rm min}$, where $R_{\rm maj}$ and $R_{\rm min}$ are the lengths of the semi-major and semi-minor axes of the structure, respectively, as shown in Fig. \ref{fig:filaments}.

In order to investigate the gravitational stability of the substructures, we assume a model of isothermal cylinder and compute the mass per unit length as

\begin{equation}\label{eq:Mline}
M_{\rm line} = \frac{M_f}{2R/\sqrt{\alpha}},
\end{equation}
where $M_f$ is the mass and $\alpha = R_{\rm min} / R_{\rm maj}$ the ellipticity of the substructure. The virial mass per unit length is defined following \citet{Fiege2000},

\begin{equation}\label{eq:Mvir}
M_{\rm line, vir} = \frac{2\sigma^2}{G}.
\end{equation}

Listed in Table \ref{tab:filament_properties} are the average hydrogen column density, the average density, the average velocity dispersion, the mass and the $M_{\rm line}$ (Eq. (\ref{eq:Mline})) and $M_{\rm line,vir}$ (Eq. (\ref{eq:Mvir})) values of the substructures. The densities and the masses are derived from the continuum observations. The values are computed over the areas shown in Fig. \ref{fig:filaments}, assuming a distance of 620 pc. For consistency with the clump parameters in Table \ref{tab:clump_masses}, the average velocity dispersion is computed from the $\rm ^{13} CO$ data.

\begin{table*}
\caption{The properties of the selected substructures identified in Fig. \ref{fig:filaments}, assuming a distance of 620 pc.}
\label{tab:filament_properties}
\centering
\begin{tabular}{c c c c c c c}
\hline\hline
& & & & & & \\
 & $N(\rm H_2)$\tablefootmark{\rm (1)} & $n(\rm H_2)$\tablefootmark{\rm (1)} & $\sigma_{\rm ^{13} CO}$ & $M_s$\tablefootmark{\rm (1)} & $M_{\rm line}$\tablefootmark{\rm (1)} & $M_{\rm line,vir}$ \\
Substructure & $\rm ( 10^{21}$ $\rm cm^{-2})$  & $\rm ( 10^3$ $\rm cm^{-3})$ & (km $\rm s^{-1}$) & $(M_{\odot})$ & $(M_{\odot}$/pc) & $(M_{\odot}$/pc) \\
\hline
& & & & & & \\

1  & 7.99 $\pm$ 0.27 & 6.40 $\pm$ 0.07 & 0.69 $\pm$ 0.02 &  76.2 $\pm$  3.5 &  94.2 $\pm$ 2.9 &  86.2 $\pm$ 3.4\\
2  & 11.2 $\pm$ 0.58 & 12.3 $\pm$ 0.11 & 0.71 $\pm$ 0.03 &  57.4 $\pm$  2.3 &  96.9 $\pm$ 4.0 & 221.1 $\pm$ 4.4\\
3  & 7.91 $\pm$ 0.28 & 7.06 $\pm$ 0.08 & 0.69 $\pm$ 0.02 &  60.7 $\pm$  4.8 &  83.6 $\pm$ 6.9 & 193.7 $\pm$ 4.7\\
4  & 7.31 $\pm$ 0.49 & 8.42 $\pm$ 0.10 & 0.73 $\pm$ 0.02 &  33.6 $\pm$  3.3 &  59.8 $\pm$ 3.9 & 221.9 $\pm$ 3.6\\
5  & 3.66 $\pm$ 0.14 & 3.51 $\pm$ 0.07 & 0.43 $\pm$ 0.01 &  24.3 $\pm$  4.4 &  36.0 $\pm$ 2.9 &  84.7 $\pm$ 3.9\\
6  & 6.50 $\pm$ 0.36 & 5.40 $\pm$ 0.06 & 0.82 $\pm$ 0.02 &  57.6 $\pm$  4.1 &  73.8 $\pm$ 3.6 & 283.3 $\pm$ 4.9\\
7  & 6.94 $\pm$ 0.27 & 4.16 $\pm$ 0.08 & 0.79 $\pm$ 0.03 & 118.2 $\pm$  4.6 & 109.3 $\pm$ 4.3 & 248.7 $\pm$ 7.2\\
8  & 7.50 $\pm$ 0.32 & 3.74 $\pm$ 0.04 & 0.58 $\pm$ 0.02 & 183.9 $\pm$  7.0 & 141.6 $\pm$ 5.4 & 178.5 $\pm$ 8.5\\
9  & 6.09 $\pm$ 0.62 & 5.11 $\pm$ 0.03 & 0.56 $\pm$ 0.01 &  52.9 $\pm$  4.2 &  68.5 $\pm$ 2.9 & 142.3 $\pm$ 2.8\\
10 & 9.65 $\pm$ 0.29 & 6.09 $\pm$ 0.05 & 0.58 $\pm$ 0.03 & 148.0 $\pm$  9.3 & 144.2 $\pm$ 5.3 & 125.9 $\pm$ 7.2\\
\hline

& & & & & & \\
\end{tabular}
\tablefoot{(1) Estimated from dust emission.}
\end{table*}

\subsection{Molecular spectra of compact sources}

\begin{figure*}
\sidecaption
\includegraphics[width=12cm]{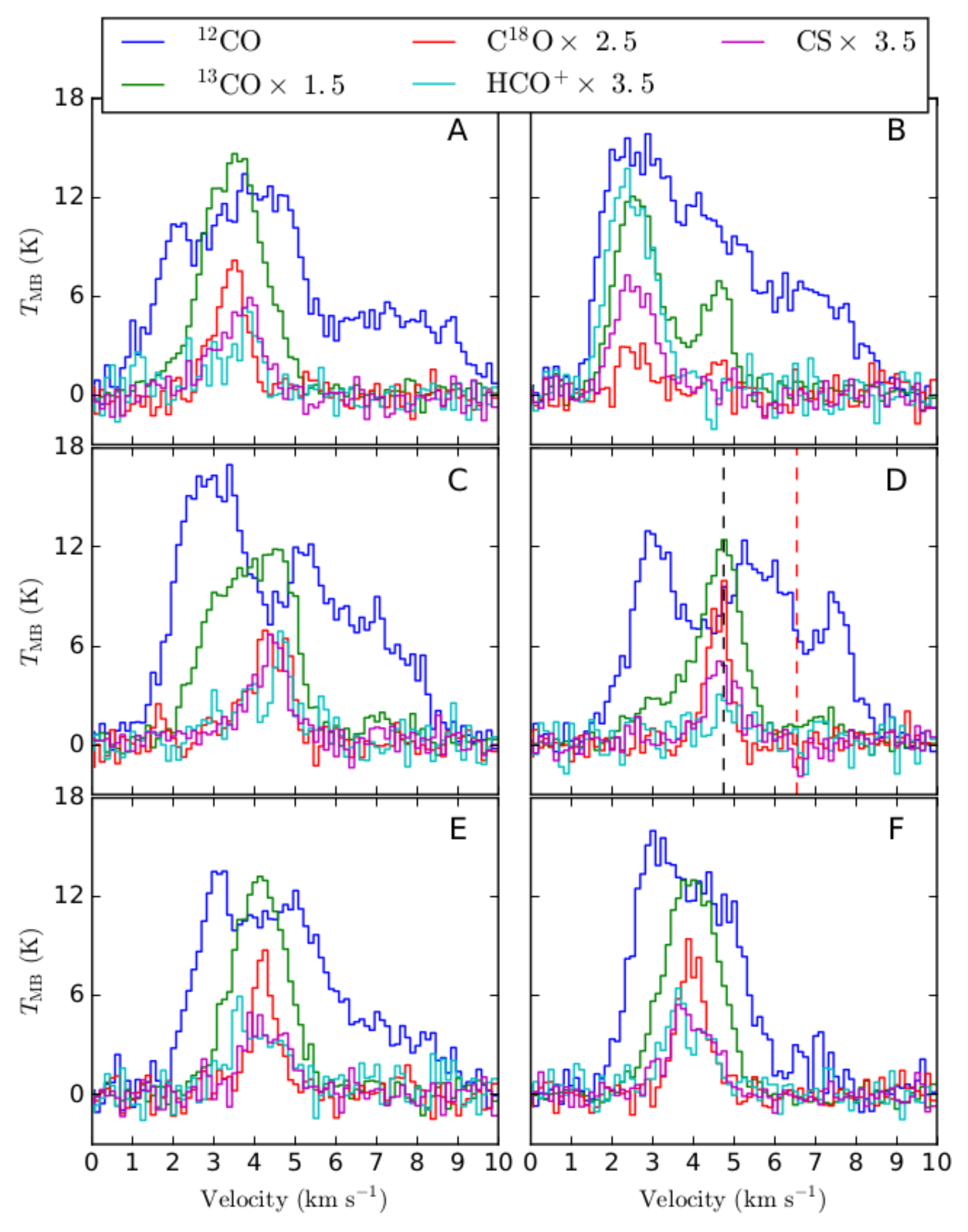} 
\caption{The spectra detected from the clumps A-F in Fig. \ref{fig:location}. The spectra are scaled by the factors shown on the top of the figure. The red and black broken lines in the panel for clump D indicate the velocity range used to calculate the mass accretion.}
\label{fig:spectra}
\end{figure*}

\begin{figure*}
\sidecaption
\includegraphics[width=12cm]{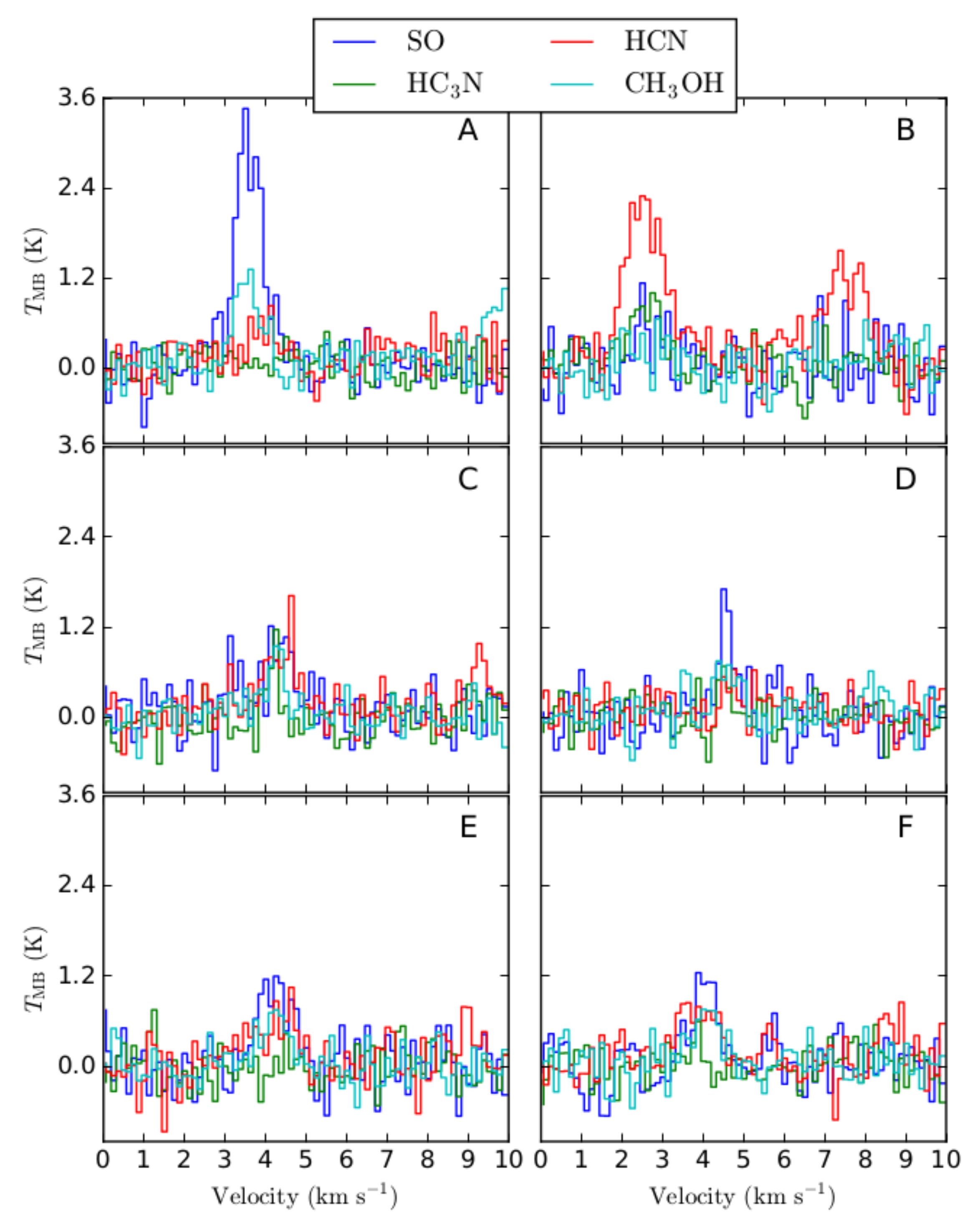} 
\caption{The spectra of the weaker lines for the clumps shown in Fig. \ref{fig:location}. Note that the secondary peak at $\sim 7$ km $\rm s^{-1}$ in the HCN spectrum of the clump B is one of the hyperfine components.}
\label{fig:spectra_rest}
\end{figure*}

For each clump, we have computed average spectra for each detected molecular line. The spectra were computed by taking an average over $50'' \times 50''$, centred at the peak of the column density and averaging over a velocity interval of 0.125 km $\rm s^{-1}$. The spectra are shown in Figs. \ref{fig:spectra} and \ref{fig:spectra_rest}. The red and black dashed lines in Fig. \ref{fig:spectra} show the velocity range used for mass accretion calculation in Section 5.

The peak of the emission is seen in the range $[3,5]$ km $\rm s^{-1}$ except for the clump B where the peak is at $\sim 2.5$ km $\rm s^{-1}$. For each clump, the $^{12} \rm CO$ line has broader features compared to the other molecular lines, often showing two or even three peaks and strong asymmetry. The complexity of the $^{12} \rm CO$ line may arise from several velocity components along the line-of-sight as can be seen in Fig. \ref{fig:cartoon}, and/or from self-absorption. The blue-shifted part of $^{13} \rm CO$ and $\rm C^{18}O$ spectra follow roughly the same shape as $^{12} \rm CO$, although for most clumps the peak of $^{13} \rm CO$ and $\rm C^{18}O$ emission is between the two peaks of the $^{12} \rm CO$ spectra. In Fig. \ref{fig:spectra}, the $\rm HCO^{+}$ and CS lines show blue asymmetric profiles (negative velocity shift by $\sim$0.5 km $\rm s^{-1}$ compared to the CO lines) for clumps E and F, a possible indication of infall. For clump A the $\rm HCO^{+}$ and CS lines have a positive velocity shift. For the remaining clumps, the $\rm HCO^{+}$ and CS have velocities consistent with the rarer CO isotopomers.

The spectra taken from the clump B show a clear double-peaked profile in $^{13} \rm CO$ and possibly in $\rm C^{18}O$. The strongest $\rm HCO^{+}$, HCN and CS emission is also seen at clump B, an indication of a different chemical composition compared to the other clumps. The HCN emission is also detected at clumps A, C, E, and F (see Fig. \ref{fig:spectra_rest}). We performed a hyperfine fit to the spectra, but obtained meaningful results only for the clump B due to low S/N ratio (see Fig. \ref{fig:HCN}). The inferred excitation temperature of HCN and the optical depth of the emission line are 6.0 K and 2.0, respectively.

The $^{13} \rm CO$ spectra taken at clumps C and D show a clear tail towards lower velocities, indicating a possible outflow or an extended foreground (or background) layer towards both clumps. The clumps A and D also show the strongest SO emission (see Fig. \ref{fig:spectra_rest}).

The $^{12} \rm CO$ velocity maps from both Osaka and Nobeyama observations show a distinct velocity component in the range $[6.5,9.5]$ km $\rm s^{-1}$, elongating orthogonal to the main filament (see Fig. \ref{fig:12CO_cont_osaka}). To study the kinematics of the component, we have extracted spectra from positions L1 and L2 (see Fig. \ref{fig:location}), and displayed them in Fig. \ref{fig:third_12CO}. The spectra show a complex velocity structure, but the strongest emission is still seen in the range of $3-4$ km $\rm s^{-1}$, with secondary maximum between 6 and 8 km $\rm s^{-1}$.

\subsection{Fragmentation and nonthermal motions}

\begin{figure}
\centering
\resizebox{\hsize}{!}{\includegraphics[width=17cm]{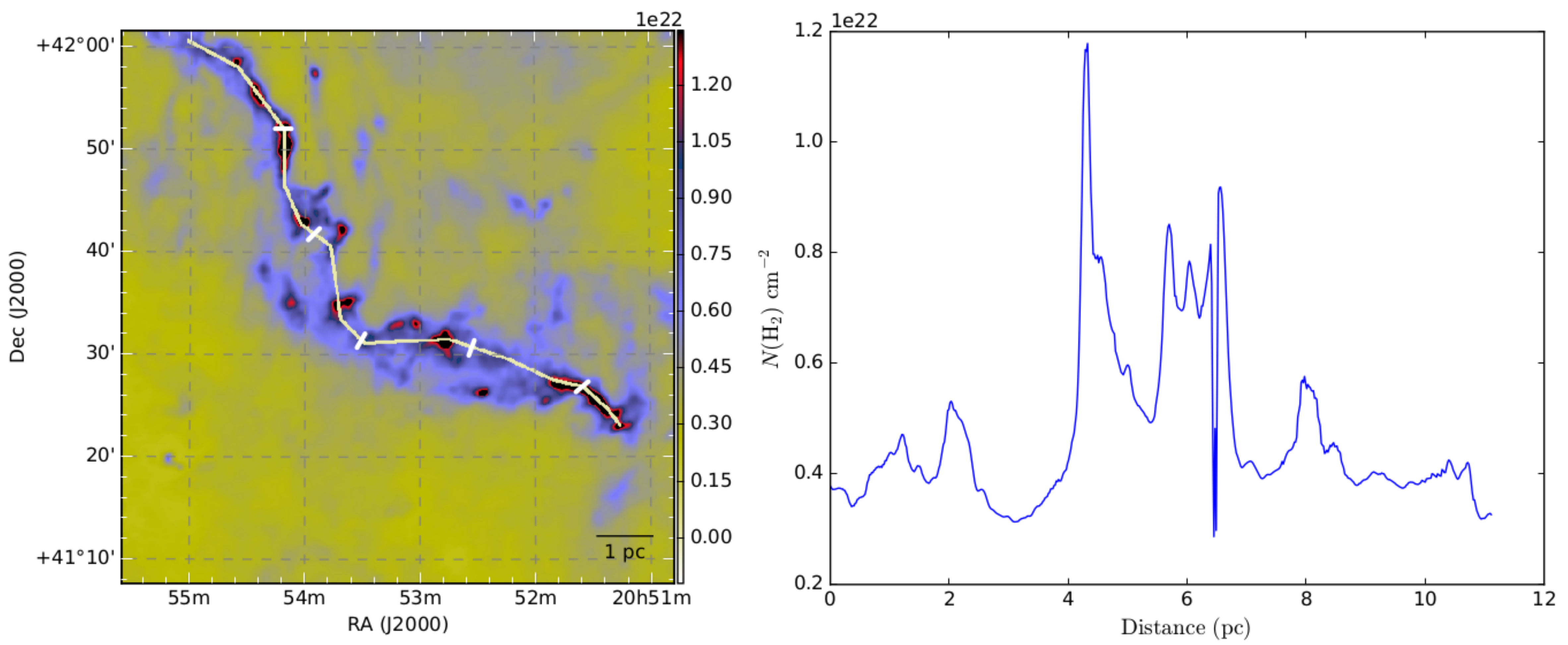}} 
\caption{Shown in the right panel is the average hydrogen column density following the white line in the left panel. The x-axis of the right panel shows the distance along the white line, starting from the upper-left corner and the white ticks on the line show steps of 2 pc. The column density values are computed as an average over $3 \times 3$ pixels. The colour map of the left panel corresponds to the hydrogen column density derived from the dust observations.}
\label{fig:density_spectra}
\end{figure}

The $Herschel$ observations show that the cold filament is fragmented to several substructures. In order to estimate a fragmentation scale length along the cold filament, we have computed an average hydrogen column density following the ridge of the cold filament. Shown in the right panel of Fig. \ref{fig:density_spectra} is the density distribution along the white line in the left panel starting from the upper left corner. There seems to be two fragmentation scales, a small scale fragmentation between individual clumps of $\sim 0.25$ pc, and a large scale fragmentation at $\sim 0.8$ pc.
The small scale fragmentation length is close to the Jeans' length \citep{Chandrasekhar_Fermi1953}

\begin{equation}
\lambda_{\rm J} = \sqrt{\frac{\pi c_s^2}{G \rho}},
\end{equation}
where $c_s$ is the sound speed, $G$ is the gravitational constant and $\rho$ is the density. Inserting $c_s = 0.21$ km s$^{-1}$ (assuming a temperature of 15 K) and $\rho = 2.16 \times 10^3$ cm$^3$ (average over the values in Table \ref{tab:filament_properties}), results in $\lambda_{\rm J} = 0.3$ pc. 

On the other hand, at larger scales the fragmentation length follows more closely the fragmentation model of infinitely long cylinder \citep{Inutsuka_Miyama1992, Jackson2010}, where the separation length $\lambda_{\rm max}$ depends on the filament scale hight $H = \sqrt{\pi c_s^2 / G \rho}$ and the radius $R$ of the filament. If $R >> H$, the fragmentation length is $\lambda_{\rm max} = 22H$. Using the vales above for $c_s$ and $\rho$, the separation length is $\lambda_{\rm max} = 1.05$ pc.

The nonthermal motions of the gas can be associated to turbulence or core formation. To study the nonthermal contribution to the line width, the thermal component has to be subtracted. We assume that the two components are independent and define the nonthermal velocity dispersion as

\begin{equation}
\sigma_{\rm nt} = \sqrt{\frac{\Delta V^2}{8 \ln{2}}-\frac{k_{\rm b} T_{\rm kin}}{m}}.
\end{equation}
The nonthermal velocity dispersion can be compared with the sound speed of the gas $c_s$, for which we have used a value of 0.21 km $\rm s^{-1}$, corresponding to a gas temperature of 15 K.

\begin{figure}
\centering
\resizebox{\hsize}{!}{\includegraphics[width=17cm]{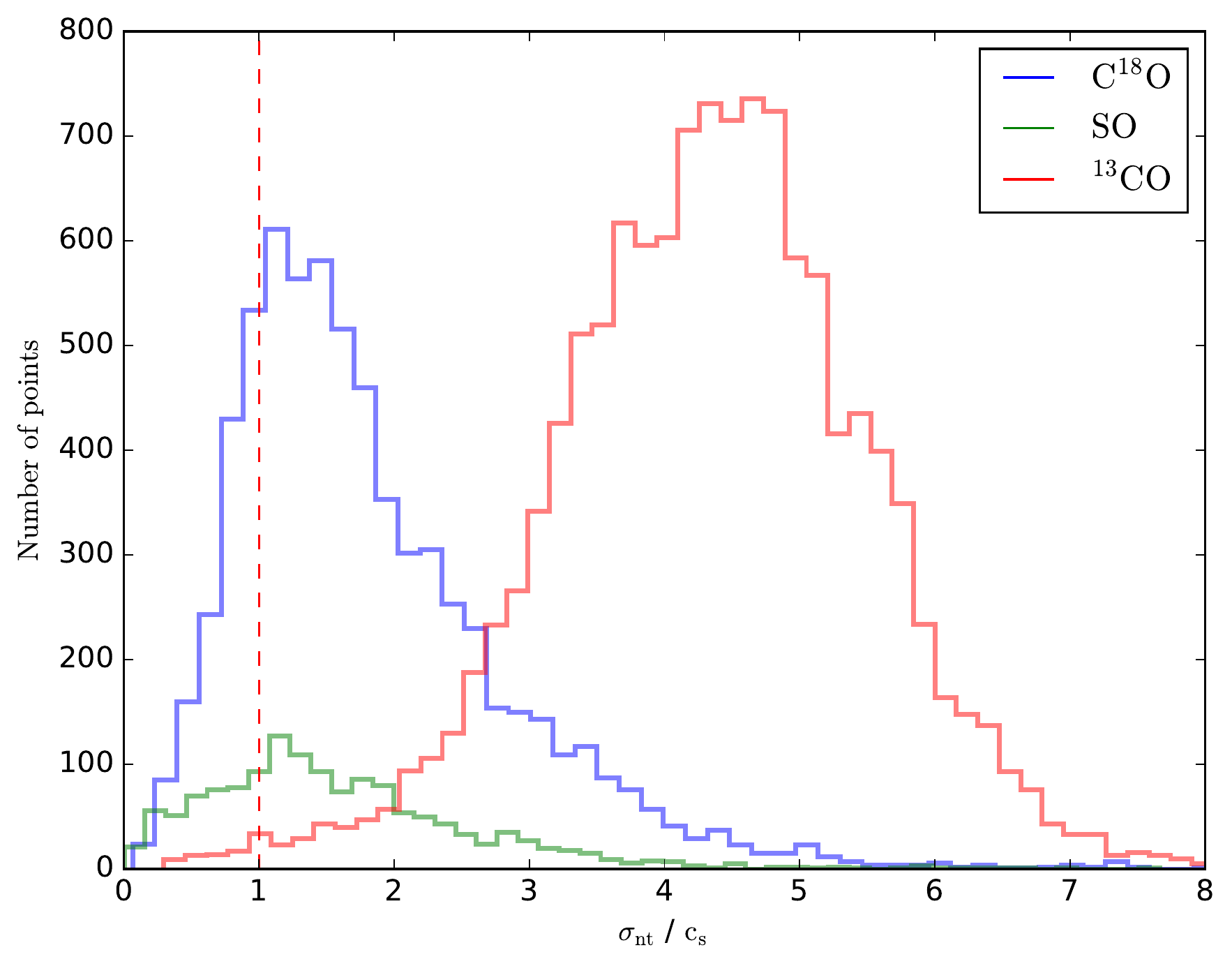}} 
\caption{The ratio of nonthermal velocity dispersion ($\sigma_{\rm nt}$) over sound speed ($c_s$) for SO, $\rm ^{13} CO_{\rm NO}$, and $\rm C^{18}O_{\rm NO}$. The red line represents the limit $\sigma_{\rm nt} / c_s$ $=$ 1.0.}
\label{fig:motion_hist}
\end{figure}

Shown in Fig. \ref{fig:motion_hist} are the histograms of the velocity dispersion of SO, $\rm ^{13} CO$, and $\rm C^{18}O$. For the histograms, only the pixels with S/N > 4.0, computed from the total integrated intensity,  were included. Only a few points have subsonic velocity dispersion, and an overwhelming majority have supersonic motions. The subsonic regions are located close to or between the outer regions of clumps B, D, and E (Fig. \ref{fig:location}). 

The linewidth-size relationship, one of the Larson's relations \citep{Larson1981}, defines the correlation between the velocity dispersion of the gas as a function of the scale. Shown in the left panel of Fig. \ref{fig:larson} is the relation derived from the Osaka $\rm ^{13}CO$ data. The data points in the panel were computed by dividing the cold filament to length bins, starting from the north-western part of the cold filament and moving to south-east along the filament. For the first point, the length of each bin is 0.25 pc, and for each subsequent data point the length of the bins was increased by 0.25 pc, up to 9.0 pc. For each data point in the figure, the velocity dispersion is computed as an average over the bins. As seen in the left panel of the Fig. \ref{fig:larson}, the velocity dispersion as a function of the size scale is relatively uniform. However, at the scale of one parsec, the scale of individual cloud fragments, the dispersion shows large variations along the filament (see right panel of Fig. \ref{fig:larson} and Fig. \ref{fig:density_spectra}).

\section{Discussion}\label{Sec4}



\subsection{Colliding filaments}

The line observations (Figs. \ref{fig:12CO_cont} and \ref{fig:13CO_cont}) indicate that the velocity field around the central part of the filament, between clumps C and F (right panel of Fig. \ref{fig:location}), is not uniform. The large scale images from $Planck$, $WISE$ and $Herschel$ PACS instrument (see Figs. \ref{fig:warmmap} and \ref{fig:coldmap}) show a warm structure, which is seen faintly in the $Herschel$ image (Fig. \ref{fig:RGB}) as a red filamentary structure running from South-East to North-West across the cold filament. The same warm filament can be seen in both the Osaka and the Nobeyama $\rm ^{12}CO$ observations in the range of $\sim[7,9]$ km $\rm s^{-1}$, Figs. \ref{fig:12CO_cont_osaka} and \ref{fig:12CO_cont}, respectively. Furthermore, the Osaka $\rm ^{12}CO$ observations in the range of $\sim [-4,-1]$ km $\rm s^{-1}$ show a separate velocity component at the western end of the mapped region, which is not associated either to the cold or the warm filament. 

The complex velocity field as well as the temperature difference of the cold and warm filaments suggest a possibility that the two filaments are distinct structures without physical connection lying along the same line-of-sight. The distance estimates are 620 pc for the cold filament and, depending on the mask used, 300 pc or 670 pc for the warm filament. Thus, if the warm filament is truly $\sim 300$ pc closer than the cold filament, the temperature difference can be explained by differences in the local radiation fields. However, if the warm filament is at a distance similar to the cold filament, the temperature difference should represent the difference in density, i.e., the cold filament is denser than the warm filament.

Looking at the $Herschel$ optical depth map (right panel of Fig. \ref{fig:location}), there are two cavities, one between clumps B and D, and the other between clump D and E. Interestingly, the cavities seem to be aligned roughly along the warm filament (see Fig. \ref{fig:RGB}). If the distance to the warm filament is 670 pc, it would seem that the clumps D and C (possibly also clump B) are close to areas where a collision between two structures is possible. 

SO can be used as a tracer molecule for shocks and outflows \citep{desFortes1993, Jorgensen2004}, thus the SO emission from the two clumps, D and B, can indicate that the clumps have formed as a result of colliding structures. In addition, both clumps show $\rm CH_{3}OH$ emission in the same velocity interval as SO emission (Figs. \ref{fig:CH3OH0_cont} and \ref{fig:SO_cont}), which could be an indication of outflows or shocks \citep{Jorgensen2004}. The SO contours in Fig. \ref{fig:SO_cont}, seem to trace the inner edges of the two cavities, indicating a possible shock. However, for both SO and $\rm CH_{3}OH$ the S/N is low, and more sensitive observations are needed.

\begin{figure}
\centering
\resizebox{\hsize}{!}{\includegraphics[width=17cm]{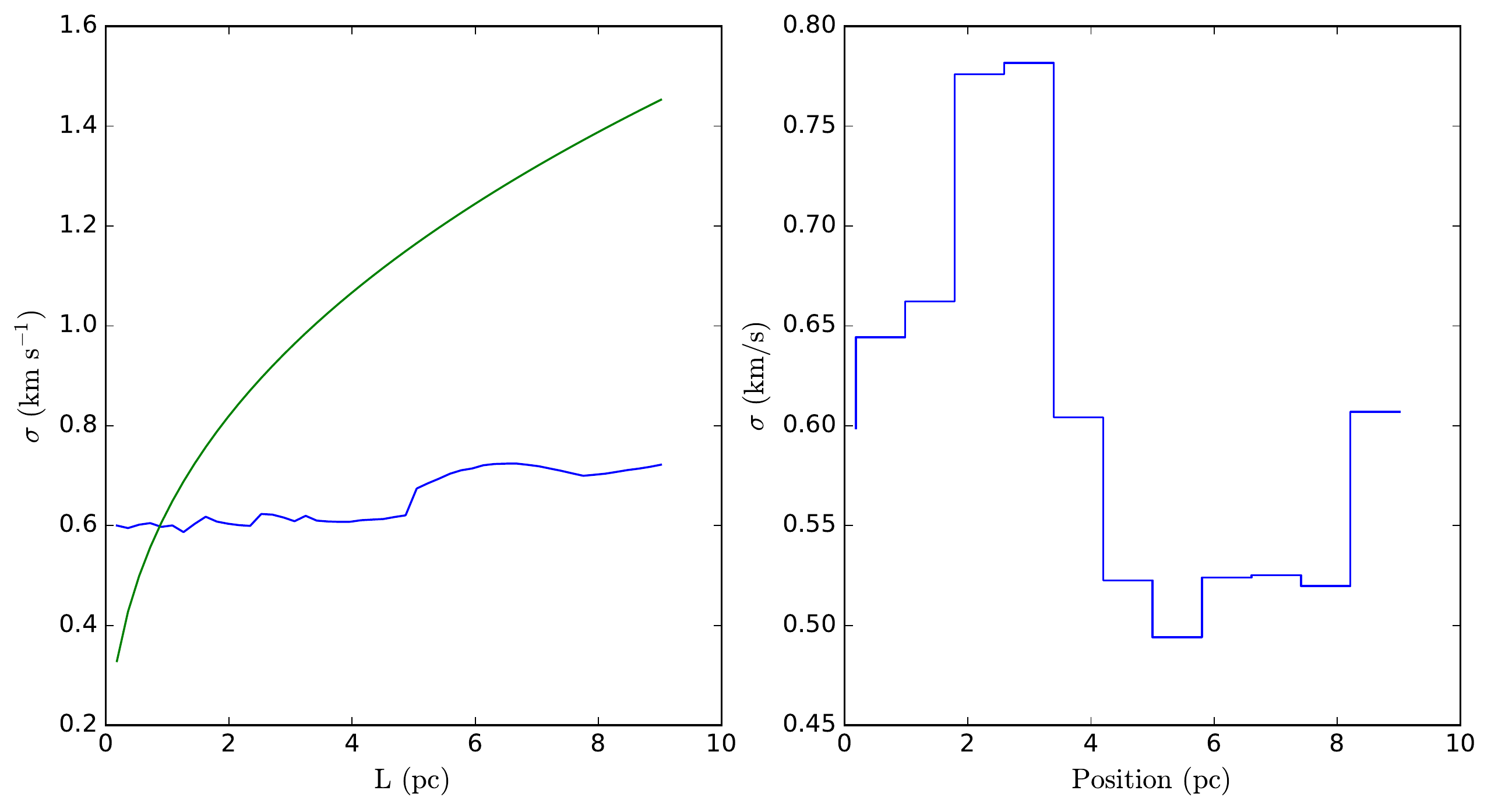}} 
\caption{Left panel: correlation of the velocity dispersion as a function of the scale, the blue line, derived using the Osaka $^{13} \rm CO$ data. The green curve is the Larson's relation  $\sigma = 0.63 \rm L^{0.38}$, where L is the scale length. Right panel: velocity dispersion as a function of position in the filament, starting from north-west and moving to south-east along the filament at one parsec scale.}
\label{fig:larson}
\end{figure}

A collision between the cold filament and the warm filament is possible if the two components are at a similar distance. In such an event, most of the clumps identified from the central part of the main filament are close to the shock front, and thus could be formed as a result of the collision. 

If the warm filament is located at $\sim300$ pc, the collisional model is untenable. Nevertheless, the $Herschel$ observations show that the central region of the cold filament is significantly broader compared to the northern and southern ends of the filament, indicating that the filament is interacting with the surrounding medium. Possible explanation could be that the entire filament is moving towards North-West with a higher relative velocity relative to the surrounding medium, or is interacting with radiation from nearby bright stars.

\subsection{Striations}

Several papers, for example \citet{Goldsmith2008} and \citet{Palmeirim2013}, have studied the relationship between striations and the parent filament. Some recent studies have shown a close connection between the striations and the large scale magnetic field \citep[e.g.][]{Malinen2016, Cox2016}. The $Herschel$ observations of G82.65-2.00 clearly show a host of faint structures around the main filament. For the purposes of this work we will identify all faint and well defined elongated structures as striations.
 
The striations raise a question about the dust properties, as the striations are faint but clearly appear in blue in Fig. \ref{fig:RGB}, and thus are clearly separated from the warm background. The average column density of the striation S1, listed in Table \ref{tab:filament_properties}, is lower by a factor of $\sim$ 2 compared to the rest of the filament. However, the dust seems to be almost as cold in the striations as in the cold filament and clearly colder than in the general background that mainly corresponds to the warm filament visible in, for example, the 12$\mu$m image (see Fig. \ref{fig:warmmap}). Thus, either the striations are not connected to the cold filament and are in a different radiation field, or the properties of the dust grains (mainly the size of the grains) have changed.

However, the $\rm ^{13}CO$ and $\rm C^{18}O$ observations from Nobeyama connect one of the striations, S1, to the cold filament. Indeed, the $\rm ^{13}CO$ and $\rm C^{18}O$ (Figs. \ref{fig:13CO_cont} and \ref{fig:C18O_cont}) contours around $\sim 5$ km $\rm s^{-1}$ trace the striation S1 and the two clumps, C and D, in the central part of the filament. Furthermore, the PV cut along the striation S1 shows a continuous velocity structure. Thus, it is likely that the striation is connected to the main filament and might be an indication of matter accretion on the filament. On the other hand, because of the active star formation or because of some external forces, and if the cold colour temperature of the striations is a sign of larger dust grains, this might also be material stripped away from the filament. 

We can derive a rough estimate for the potential mass accretion from the striation S1 onto the clump D, by assuming a simple relation for a core accreting mass along a cylindrical structure

\begin{equation}
\dot{M} =  \frac{M_s \hat{V}}{L_{s}},
\end{equation}
where $M_s$ is the mass of the striation, $L_s$ is the length of the striation, and $\hat{V}$ is the velocity. As the value of $\hat{V}$, we use the difference between the central velocity of the $^{13} \rm CO$ line (black dashed line in Fig. \ref{fig:spectra}) and the extent of the red lobe of $^{12} \rm CO$ line (red dashed line in Fig. \ref{fig:spectra}). Inserting $L_s$ = 2.17 pc, $\rm \hat{V} = 1.7$ km $\rm s^{-1}$, and $M$ = 36.0 $M_{\odot}$, we obtain an estimate for the mass accretion rate of $\dot{M}$ = $2.2 \times 10^{-6}$ $M_{\odot} / \rm year$.

In a recent study \citet{LopezSepulcre2010} analysed a selection of 48 high mass molecular clumps, 10 of which were inferred as infall candidates. The mass accretion rates for the 10 sources were derived from the outflow mass loss rates (for a review of the method see \citet{Beuther2002} and \citet{LopezSepulcre2009}) resulting in accretion rates of $\sim 10^{-5}$ $M_{\odot} / \rm year$. Similarly, \citet{Palmeirim2013} studied the Taurus B211/3 filament identifying several striations that seem to be connected to the main filament. \citet{Palmeirim2013} explored several methods to derive the infall velocity, deriving an accretion rates between $2.7-5.0\times 10^{-5}$ $M_{\odot} / \rm year$ from a 220 $M_{\odot}$ striation onto the main filament and adopting a velocity of $\sim 1$ km $\rm s^{-1}$.

Comparing the above numbers, our derived value is lower by an order of magnitude. However, the environments and accretion models under consideration differ significantly and because our clumps are smaller, a smaller accretion rate seems reasonable.

\subsection{Kinematics}
The $M_{\rm line}$ values for almost all of the substructures are close to $\sim 100$ $M_{\odot}/$pc, which seems to hold also for larger scales. The estimated total mass of the filament is $\sim$1200 $M_{\odot}$ and the length of the filament is $\sim$ 10 pc. Thus, the resulting $M_{\rm line}$ for the main filament is $\sim 120$ $M_{\odot}/$pc. Only three of the substructures (Nos. 1, 8, and 10 in Table \ref{tab:filament_properties}) have $M_{\rm line}$ greater than or close to $M_{\rm line, vir}$. The result indicates that most of the structures are not bound and will eventually disperse back to the ISM and only the three substructures would be able to form new stars. 

However, the derived virial masses might be overestimated. The velocity dispersions used for computing the $M_{\rm line, vir}$ are derived from the large scale Osaka $\rm ^{13}CO$ observations. If the $\rm ^{13}CO$ line is saturated, the derived velocity dispersion will be higher because of the broader line profile. The pixel-to-pixel correlation between the $\tau_{250}$ and the $N(\rm ^{13}CO)$, Fig. \ref{fig:comp_2D}, indicates that the $\rm ^{13}CO$ line is indeed saturated. Thus, the velocity dispersions should be estimated from the $\rm C^{18}O$ line. However, because of the lower S/N of the $\rm C^{18}O$ data and because we do not detect $\rm C^{18}O$ in the large scale observations, we use $\rm ^{13}CO$. On the other hand, because of the high column density, the derived velocity dispersion might not give a proper picture of the inner motion of the clumps or substructures. Furthermore, the lower resolution will increase the uncertainty of the derived values. An uncertainty of 5$\%$ in the line width will change the derived $M_{\rm line, vir}$ value by several $M_{\odot}$. The $M_{\rm line}$ and $M_{\rm line, vir}$ values were computed with the assumption of isothermal cylinder. The assumption is probably not an accurate description for most of the substructures as they have already formed cores as can be seen in Fig. \ref{fig:location}. 

The $N(\rm ^{13}CO)$ line was also used to estimate the virial masses of the cold clumps (Table \ref{tab:clump_masses}) with the exception that we used the higher resolution Nobeyma observations. For most clumps the estimated virial masses are larger than the clump masses, however, for the clumps we can computed a second estimate for the velocity dispersion using the $\rm C^{18}O$ line which is not showing signs of saturation. The velocity dispersions measured in $\rm C^{18}O$ are smaller as summarized in Table \ref{tab:clump_masses}, yielding smaller virial masses, indicating that most of the clumps have masses larger than or close to the virial masses. Thus it is possible that some of the substructures might be bound.

As discussed by \citet{Arzoumanian2013}, a higher velocity dispersion can be related to mass accretion. The higher velocity dispersion values we derive for substructures 4 and 7, and possibly substructure 6, might thus be related to mass accretion, as the structures seem to be connected to striations. In such a case, the higher velocity dispersion is not caused by larger scale turbulence but rather by amplification of initial velocity fluctuations via conversion from gravitational energy to kinetic energy during mass accretion \citep{Arzoumanian2013}. However, due to limited coverage of our higher resolution observations, the connection between the substructures and the striations can not be confirmed.

The recent findings of \citet{Hacar2011} and \citet{Hacar2016}, in L1517 and in Musca, show two filaments that are clearly subsonic, and, at least in Musca, do not follow the Larson's velocity dispersion-size relationship. We see, in Fig. \ref{fig:larson}, that most of the gas, at least in the central regions, of G82.65-2.00 is clearly supersonic, with $\sigma_{\rm nt} / \sigma_{\rm s} > 1.0$. However, the derived Larson's relation is almost flat, indicating that the overall velocity dispersion along the filament, even at larger scales, is relatively constant. However, at the 1 pc scale the substructure of the larger filament is clearly seen. Thus, as the filament is likely dispersing, the dispersion is occurring slowly.


\section{Conclusions}\label{Sec5}

The $Herschel$ field G82.65-2.00 is a filamentary field at a late stage of evolution, a 'debris filament'. The central region of the cold, main filament has a clearly ragged morphology, and has formed several cold cores. Furthermore, the presence of several YSOs and the relatively high velocity differences along the cold filament can indicate that the filament is dispersing to the ISM. Our main conclusions are

\begin{enumerate}

\item Although, our line observations agree well with the results of \citet{Dame2001} who inferred a distance of 1 kpc, different methods result in inconsistent distance estimates. We have derived a new distance estimate of 620 pc $\pm$ 40 pc utilizing an infrared extinction method.

\item The $\rm ^{12}CO$ maps both from Nobeyama and Osaka combined with $Herschel$ and $WISE$ observations show a separate, diffuse warm filament extending over the cold filament. Using the extinction method, we have derived an distance estimate of 300 pc or 670 pc for the warm filament, depending on the used mask, possibly placing it at a similar distance as the cold filament. Comparing the $Herschel$ images and SO velocity contours, and assuming the two filaments are at a similar distance, it is possible that the two structures are colliding. Thus, most of the cold clumps in the central region of the cold filament could have been formed as a result of the collision.

\item The $Herschel$ observations and the line observations from Nobeyama indicate that the average molecular hydrogen column density of the filament is in the range of $1.0 - 2.5 \times 10^{22}$ $\rm cm^{-2}$. We have chosen six clumps for further analysis from the central region of the cold filament. The masses of the clumps vary in the range $10 - 20$ $M_{\odot}$ and the total mass of the clumps is $\sim 70$ $M_{\odot}$. Virial masses estimated from the velocity dispersion of the $\rm C^{18}O$ emission line are close to or slightly smaller than the clump masses. The chemical composition of the clumps is uniform, except for clump B for which the molecular line spectra shows distinct variations compared to the other clumps. We have visually identified substructures of the cold filament. The substructures have a wide range of masses from $\sim 20$ $M_{\odot}$ up to $\sim 180$ $M_{\odot}$, and in total, the mass of the cold filament exceeds 1200 $M_{\odot}$. Only three of the substructures have masses close to or greater than the virial mass. However, the estimate is based on the $\rm ^{13}CO$ line, which is likely saturated in the densest regions. As in the case with the cold clumps, the derived virial masses would likely be smaller if the velocity dispersion could be estimated from the $\rm C^{18}O$ line.


\item Utilizing the Nobeyama 45m telescope we have detected several kinematic components from the field. The strongest line emission is seen in the velocity range $[3,5]$ km $\rm s^{-1}$ with other notable emission around $-4$ and 8 km $\rm s^{-1}$. The spectral lines show a complex velocity structure around the filament, and the velocity profiles of $\rm ^{12}CO$ and $\rm ^{13}CO$ along the striation S1 seem to indicate possible mass accretion from the striation on the filament. With a simple model we estimate a tentative mass accretion rate of $\dot{M}$ = $2.23 \times 10^{-6}$ $M_{\odot} / \rm year$.

\end{enumerate}

In order to study the fragmentation and possible dispersion of the filament in greater detail, more large-scale molecular line observations would be needed. A high-resolution map  for example in ammonia would be useful to determine not only the substructure kinematics but also the state of the cold clumps. Furthermore, in connection with a more detailed study of the filamentary structures, the $Planck$ polarization data could be used to investigate the relation between magnetic fields and the identified substructures.

\begin{acknowledgements}
MS and MJ acknowledge  the support of the Academy of Finland Grant No. 285769. JMa acknowledges the support of ERC-2015-STG No. 679852 RADFEEDBACK.
\end{acknowledgements}

\bibliographystyle{aa}
\bibliography{bibli_core45}
\newpage
\begin{appendix}

\section{Osaka velocity maps}

\begin{figure*}
\centering
\includegraphics[width=17cm]{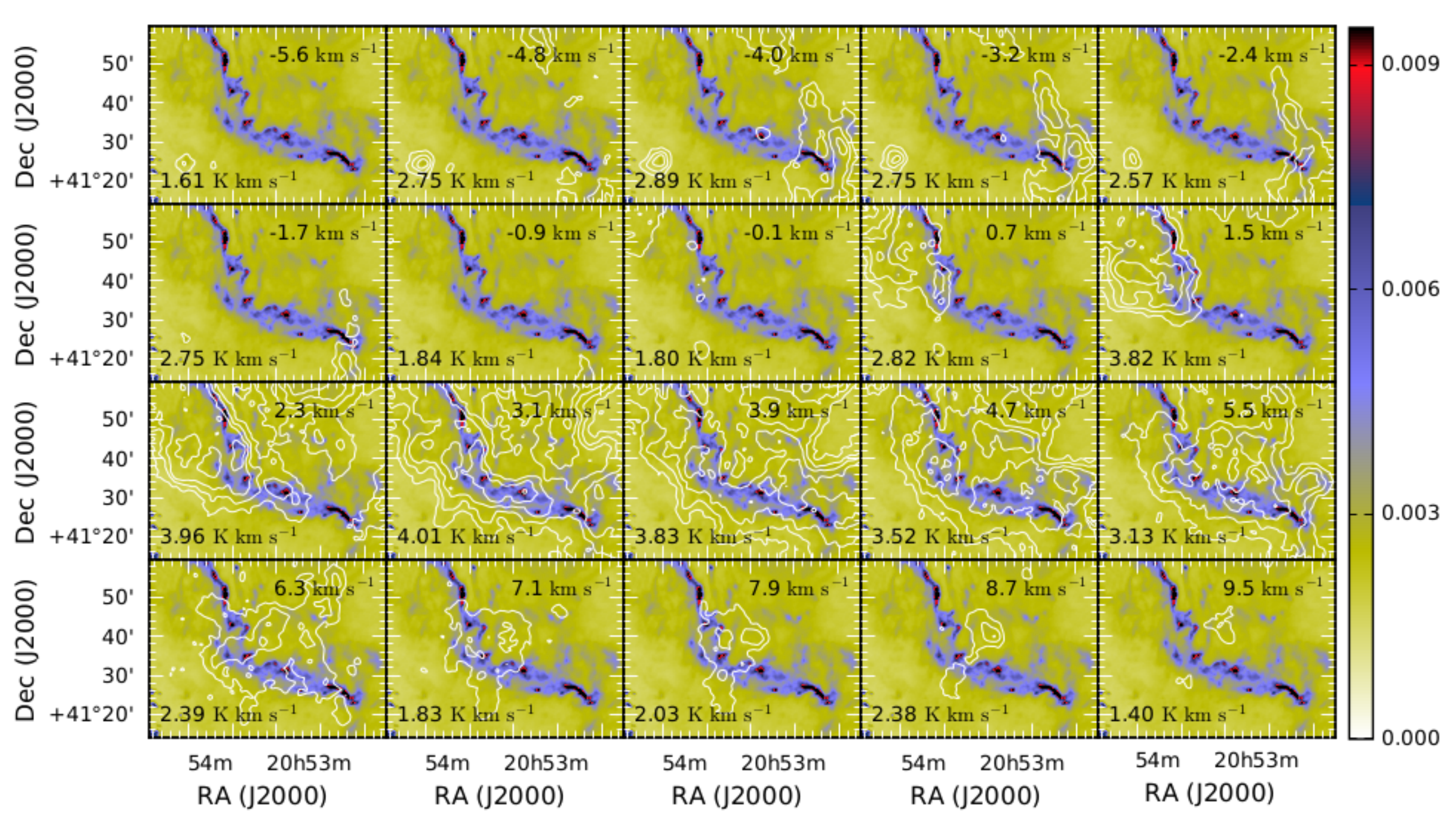} 
\caption{Osaka $\rm ^{12}CO$ line observations integrated over 0.79 km $\rm s^{-1}$ wide velocity intervals. The average noise per 0.79 km $\rm s^{-1}$ wide velocity interval is $\sigma = 0.40$  K km s$^{-1}$. The lowest contour corresponds to 3$\sigma$, and the contours increase with 3$\sigma$ steps. The numbers noted in the upper right corner of each panel correspond to the centre of the velocity bin, and the numbers in the lower left corner of each panel show the maximum intensity of the panel. The underlying colour map is the $\tau_{250}$ map.}
\label{fig:12CO_cont_osaka}
\end{figure*}

\begin{figure*}
\centering
\includegraphics[width=17cm]{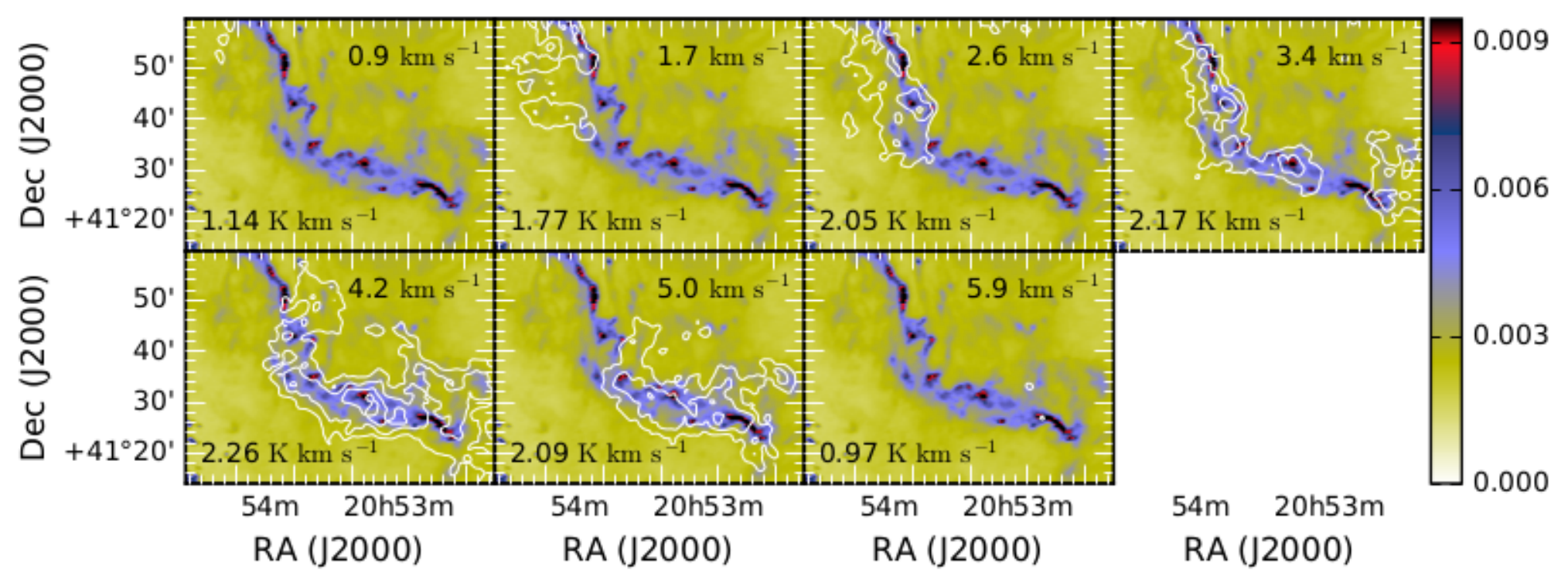} 
\caption{Same as Fig. \ref{fig:12CO_cont_osaka}, but for $\rm ^{13}CO$ and integrated over 0.83 km $\rm s^{-1}$ wide velocity intervals. The average noise per 0.83 km $\rm s^{-1}$ wide velocity interval is $\sigma = 0.26$ K km s$^{-1}$. The lowest contour corresponds to 3$\sigma$, and the contours increase with 2$\sigma$ steps.}
\label{fig:13CO_cont_osaka}
\end{figure*}

Figs. \ref{fig:12CO_cont_osaka} and \ref{fig:13CO_cont_osaka} show the velocity channel maps of $^{12} \rm CO$ and $^{13} \rm CO$ derived from the Osaka observations. For both velocity maps, the underlying colour map is the 250 $\mu$m optical depth. We do not display the $\rm C^{18}O$ channel map because of its low S/N.

\section{Nobeyama velocity maps}

\begin{figure*}
\centering
\includegraphics[width=17cm]{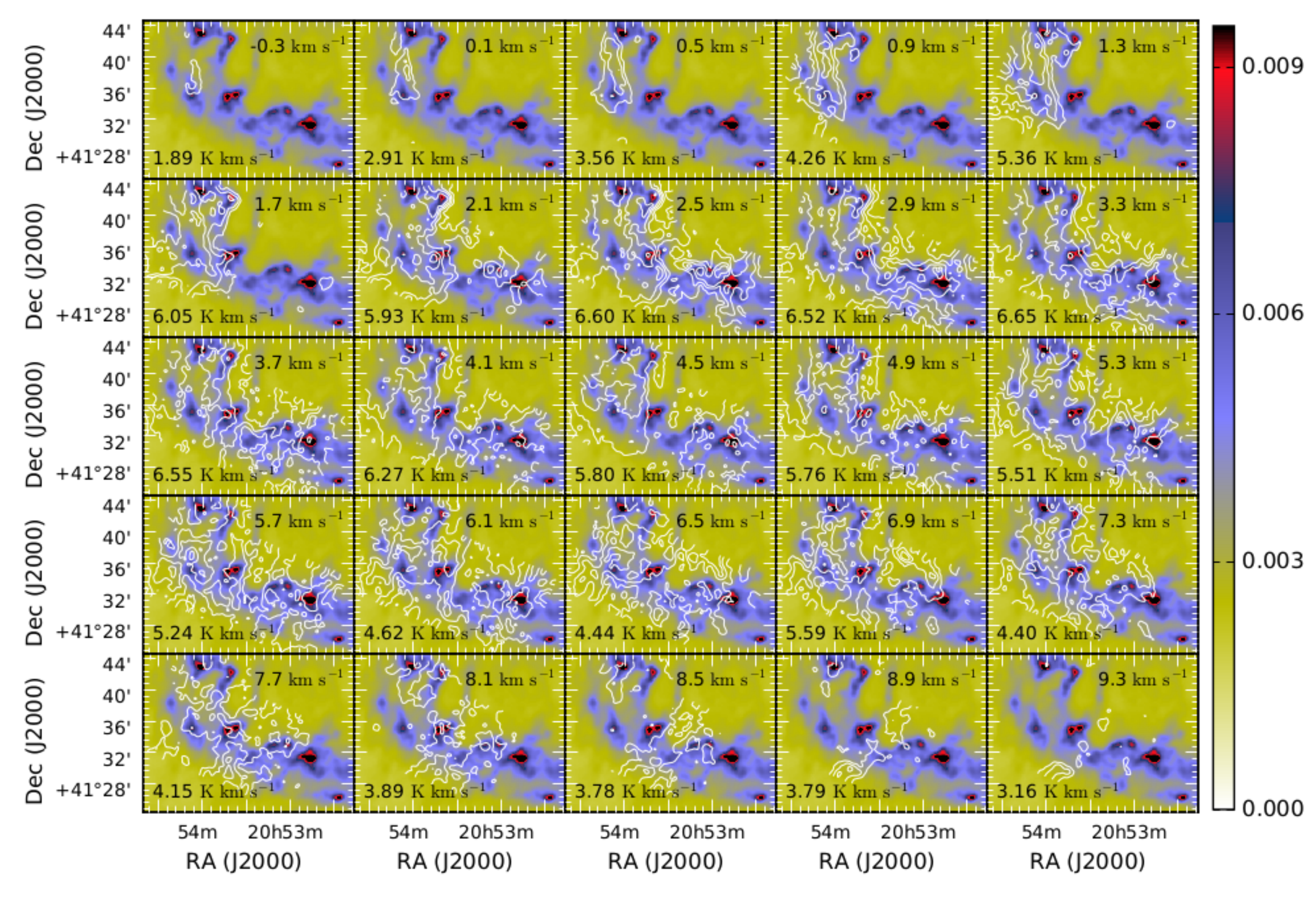} 
\caption{Integrated $\rm ^{12}CO$ line intensity $W$ in 0.4 km $\rm s^{-1}$ wide velocity intervals. The average noise per 0.4 km $\rm s^{-1}$ wide velocity interval is $\sigma = 0.12$ K km s$^{-1}$. The lowest contour corresponds to 6$\sigma$, and the contours increase with 6$\sigma$ steps. The underlying colour map is the $\tau_{250}$ map. The bin centre velocity and the maximum value of $W$ are given in the panels. }
\label{fig:12CO_cont}
\end{figure*}

\begin{figure*}
\centering
\includegraphics[width=17cm]{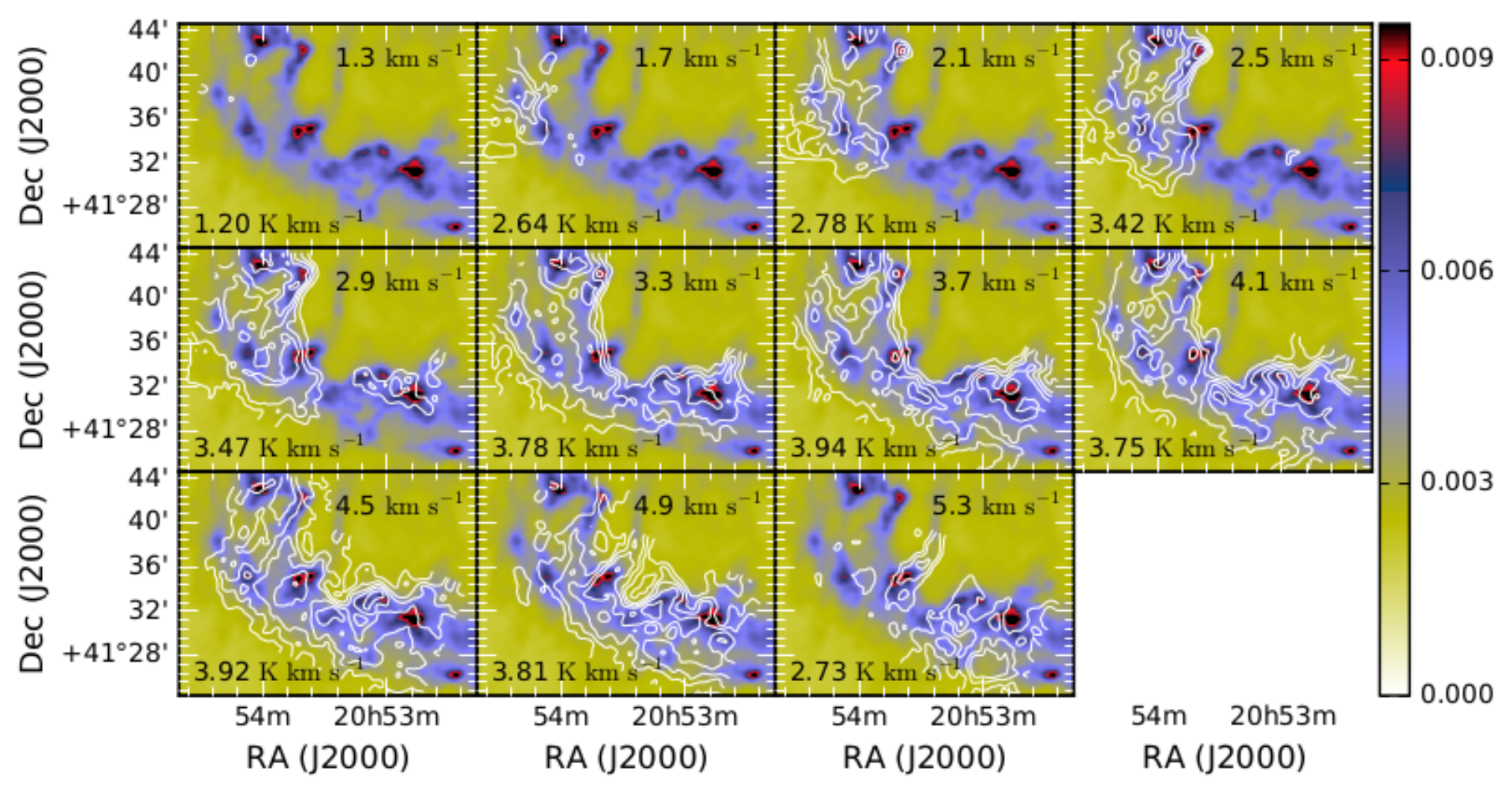} 
\caption{Same as Fig. \ref{fig:12CO_cont}, but for $\rm ^{13}CO$. The average noise per 0.4 km $\rm s^{-1}$ wide velocity interval is $\sigma = 0.15$ K km s$^{-1}$. The lowest contour corresponds to 6$\sigma$, and the contours increase with 4$\sigma$ steps.}
\label{fig:13CO_cont}
\end{figure*}

\begin{figure*}
\centering
\includegraphics[width=17cm]{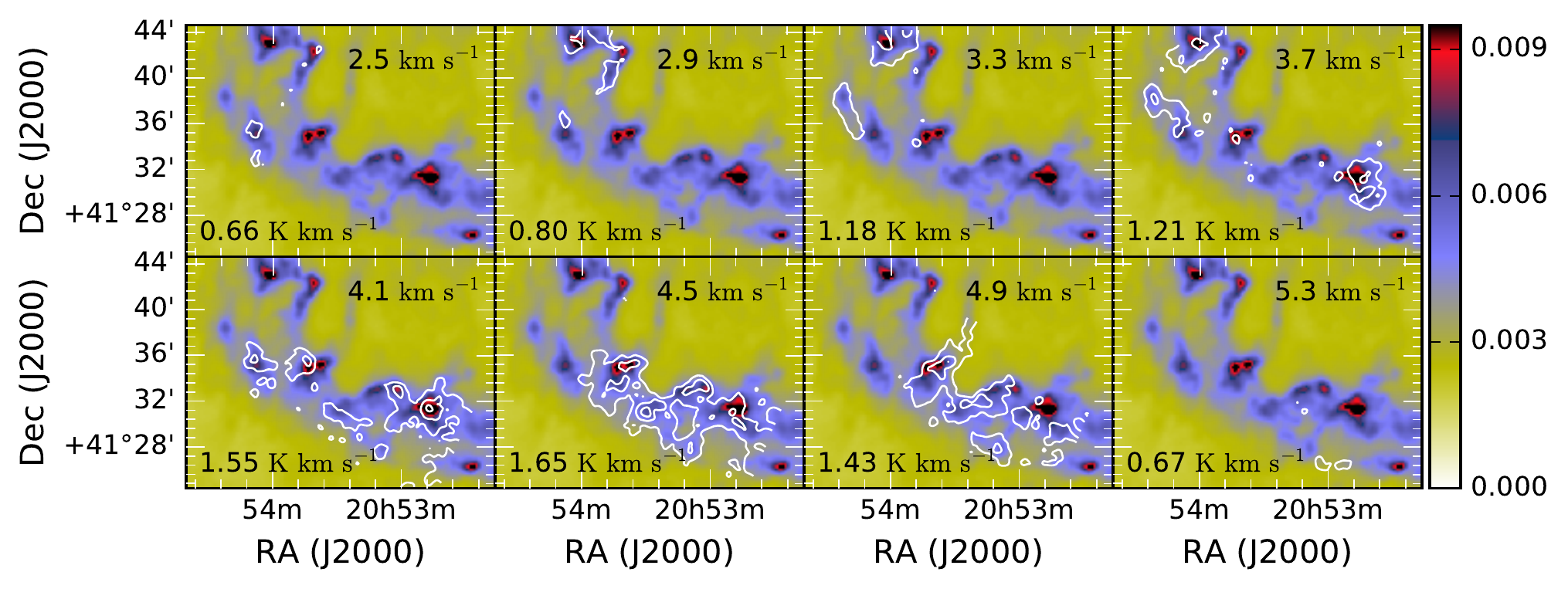} 
\caption{Same as Fig. \ref{fig:12CO_cont}, but for $\rm C^{18}O$. The average noise per 0.4 km $\rm s^{-1}$ wide velocity interval is $\sigma = 0.15$ K km s$^{-1}$. The lowest contour corresponds to 3$\sigma$, and the contours increase with 3$\sigma$ steps.}
\label{fig:C18O_cont}
\end{figure*}

\begin{figure*}
\centering
\includegraphics[width=17cm]{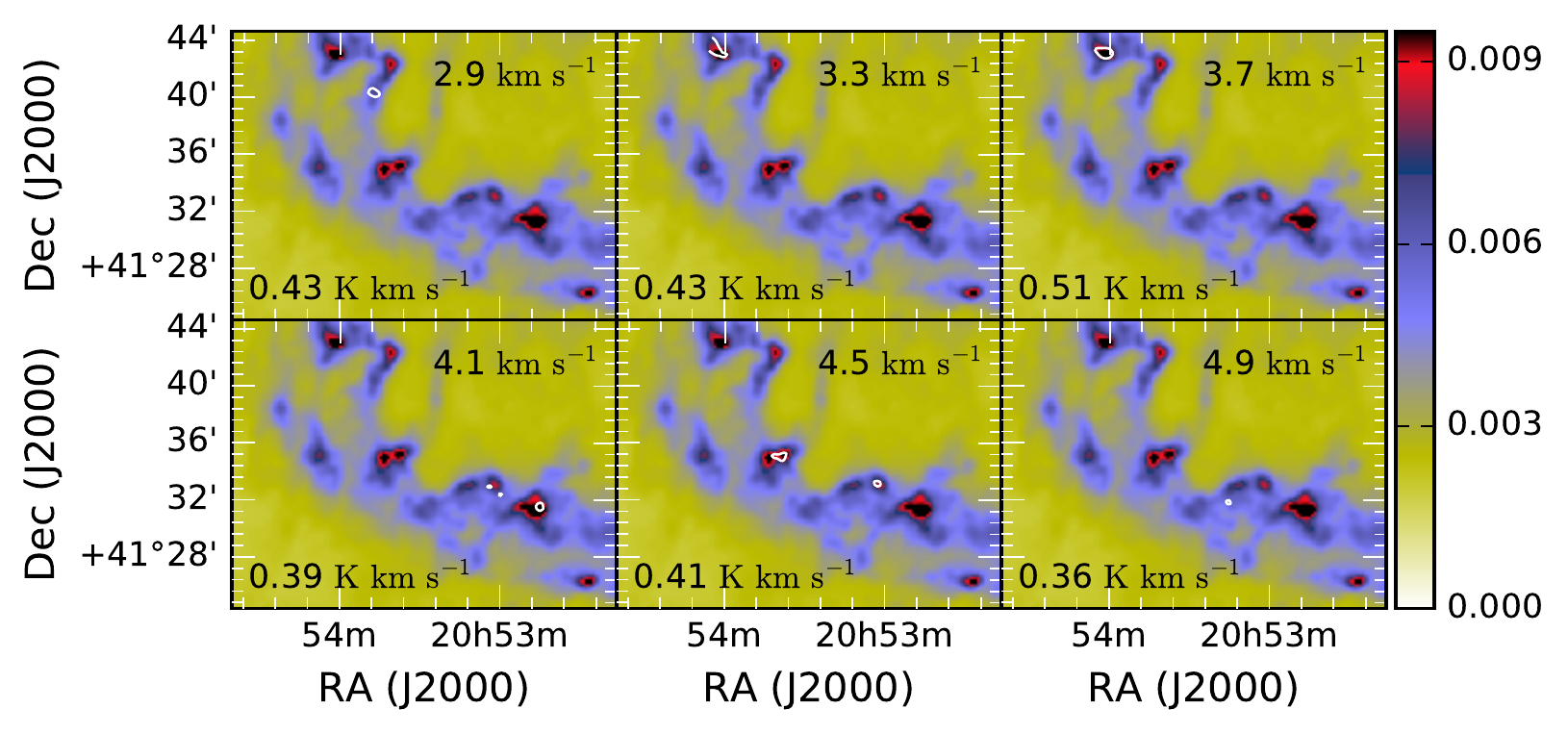} 
\caption{Same as Fig. \ref{fig:12CO_cont}, but for $\rm CH_{3}OH$. The average noise per 0.4 km $\rm s^{-1}$ wide velocity interval is $\sigma = 0.13$ K km s$^{-1}$. The contour corresponds to 3$\sigma$.}
\label{fig:CH3OH0_cont}
\end{figure*}

\begin{figure*}
\centering
\includegraphics[width=17cm]{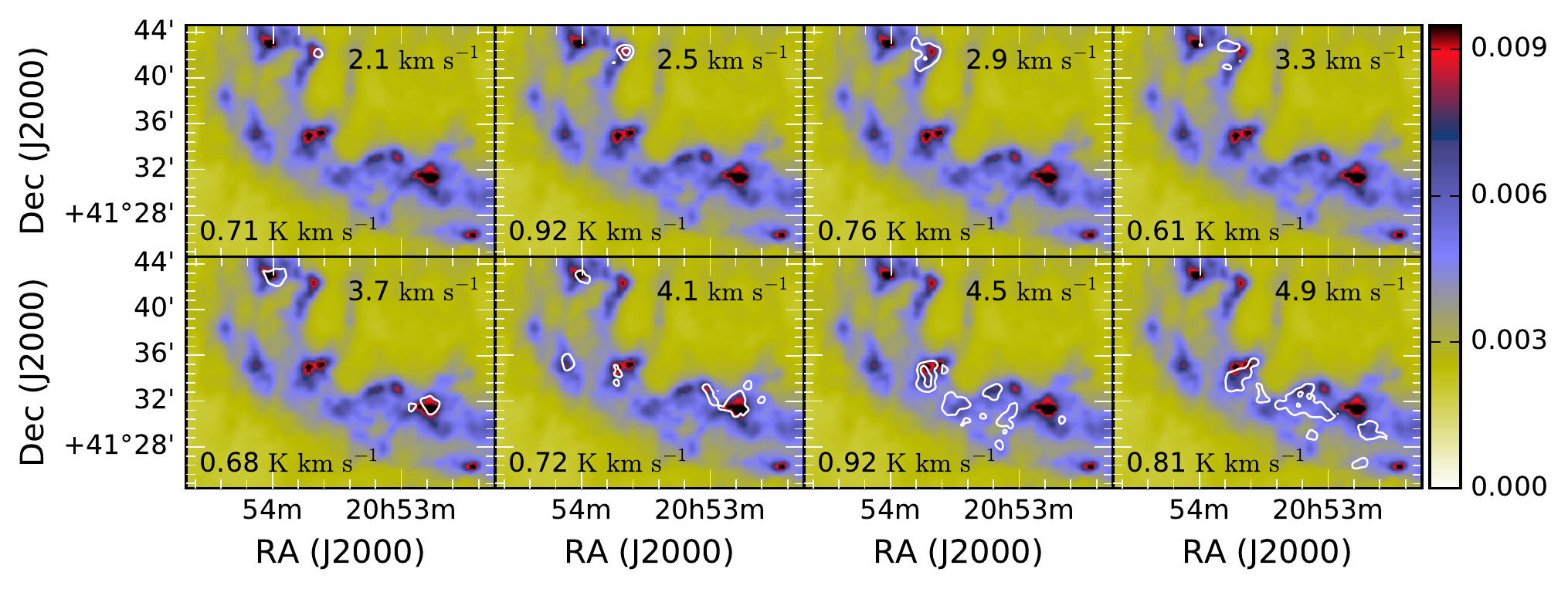} 
\caption{Same as Fig. \ref{fig:12CO_cont}, but for CS. The average noise per 0.4 km $\rm s^{-1}$ wide velocity interval is $\sigma = 0.14$ K km s$^{-1}$. The lowest contour corresponds to 3$\sigma$, and the contours increase with 2$\sigma$ steps.}
\label{fig:CS_cont}
\end{figure*}

\begin{figure*}
\centering
\includegraphics[width=17cm]{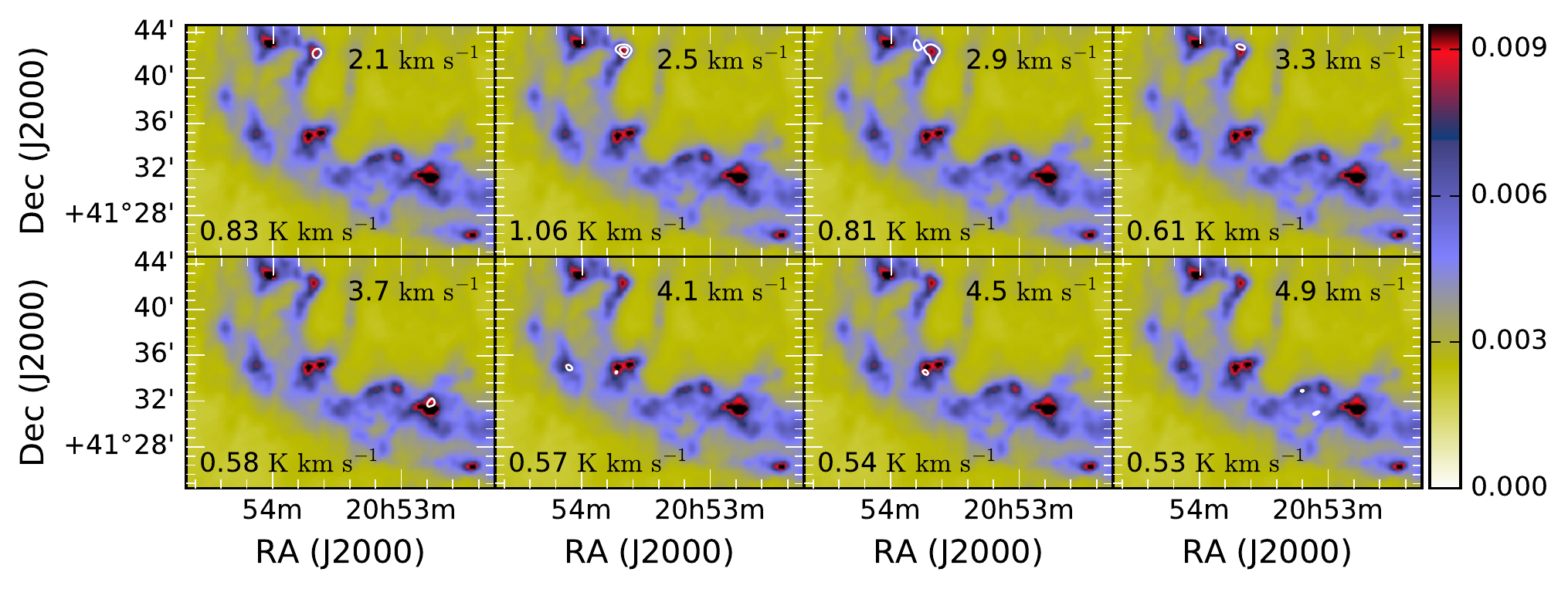} 
\caption{Same as Fig. \ref{fig:12CO_cont}, but for $\rm HCN$. The average noise per 0.4 km $\rm s^{-1}$ wide velocity interval is $\sigma = 0.16$ K km s$^{-1}$. The lowest contour corresponds to 3$\sigma$, and the contours increase with 2$\sigma$ steps.}
\label{fig:HCN_cont}
\end{figure*}

\begin{figure*}
\centering
\includegraphics[width=17cm]{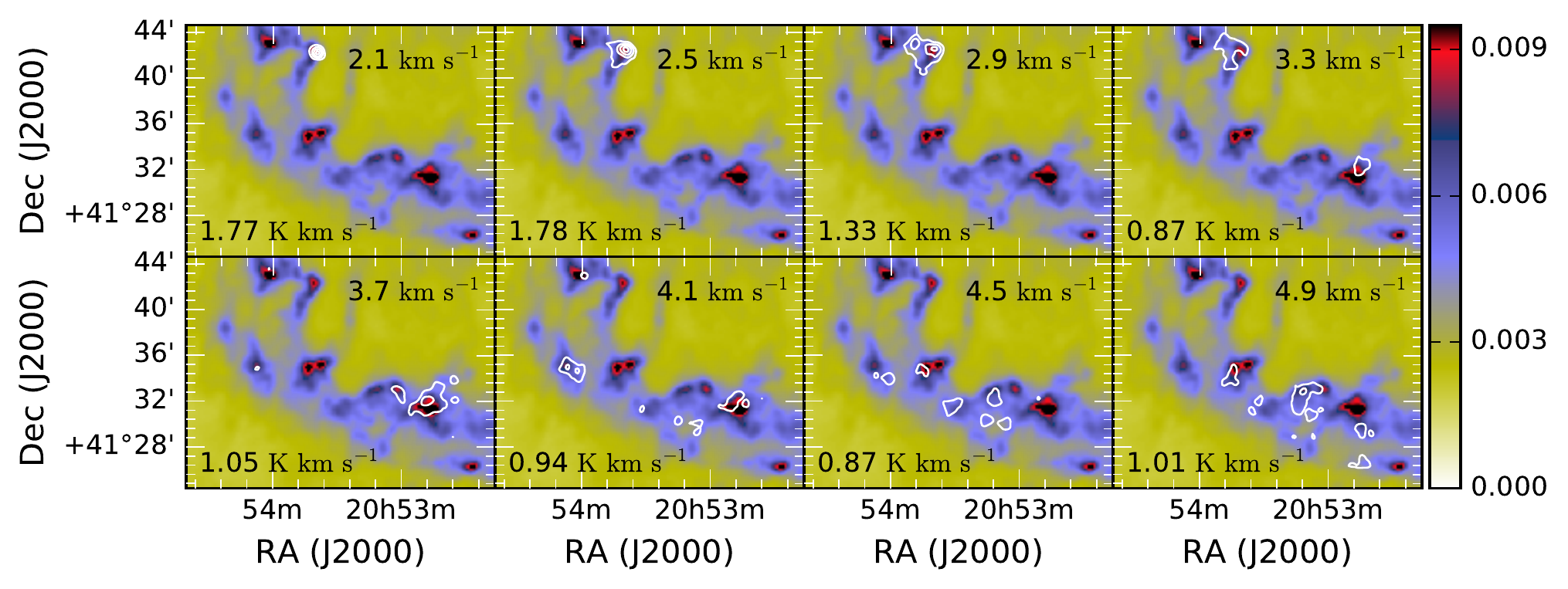} 
\caption{Same as Fig. \ref{fig:12CO_cont}, but for $\rm HCO^{+}$. The average noise per 0.4 km $\rm s^{-1}$ wide velocity interval is $\sigma = 0.17$ K km s$^{-1}$. The lowest contour corresponds to 3$\sigma$, and the contours increase with 2$\sigma$ steps.}
\label{fig:HCO_cont}
\end{figure*}

\begin{figure*}
\centering
\includegraphics[width=17cm]{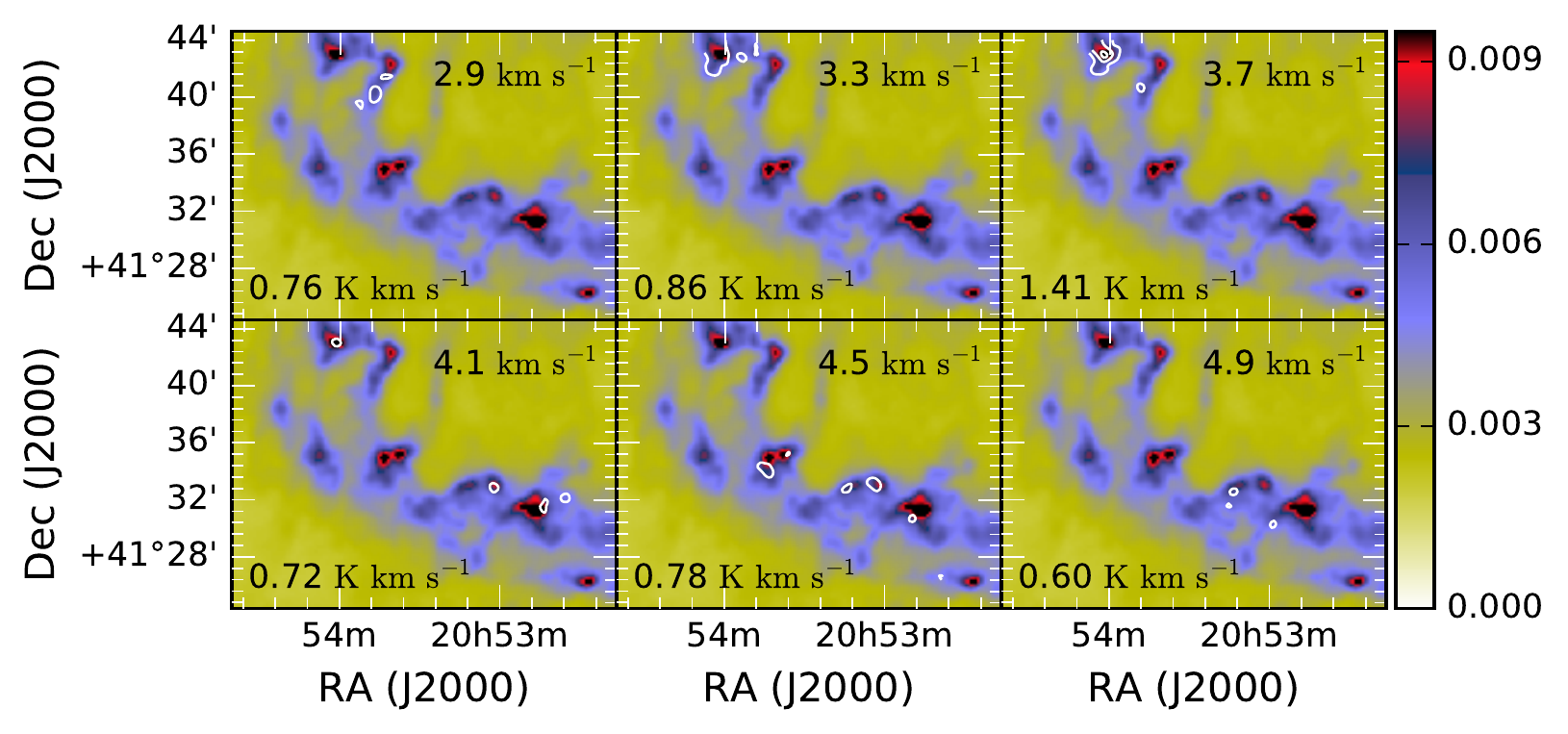} 
\caption{Same as Fig. \ref{fig:12CO_cont}, but for $\rm SO$. The average noise per 0.4 km $\rm s^{-1}$ wide velocity interval is $\sigma = 0.19$ K km s$^{-1}$. The lowest contour corresponds to 3$\sigma$, and the contours increase with 2$\sigma$ steps.}
\label{fig:SO_cont}
\end{figure*}

Figs. \ref{fig:12CO_cont} to \ref{fig:SO_cont} show the velocity channel maps for all of the molecular emission lines detected from the Nobeyama observations. In all of the figures, the underlying colour map is the 250 $\mu$m optical depth.

\begin{figure}[!h]
\centering
\resizebox{\hsize}{!}{\includegraphics[width=17cm]{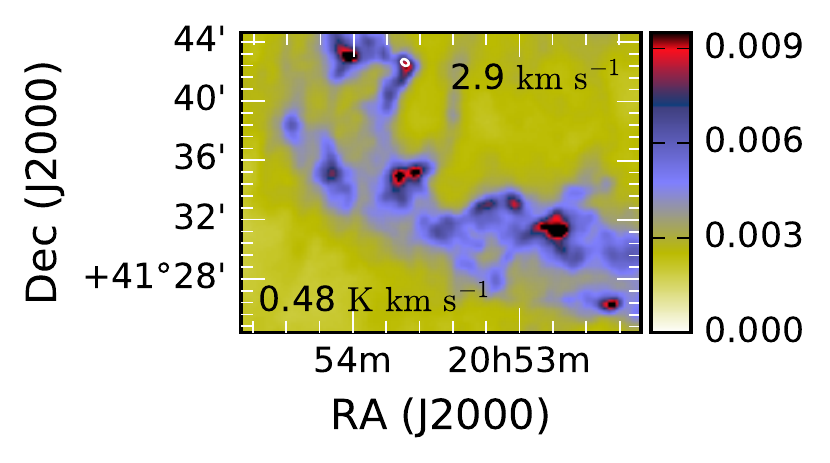}} 
\caption{Same as Fig. \ref{fig:12CO_cont}, but for $\rm HC_{3}N$. The average noise per 0.4 km $\rm s^{-1}$ wide velocity interval is $\sigma = 0.15$ K km s$^{-1}$. The contour corresponds to 3$\sigma$.}
\label{fig:HC3N_cont}
\end{figure}

\section{HCN hyperfine fitting}

\begin{figure}[!h]
\centering
\resizebox{\hsize}{!}{\includegraphics[width=17cm]{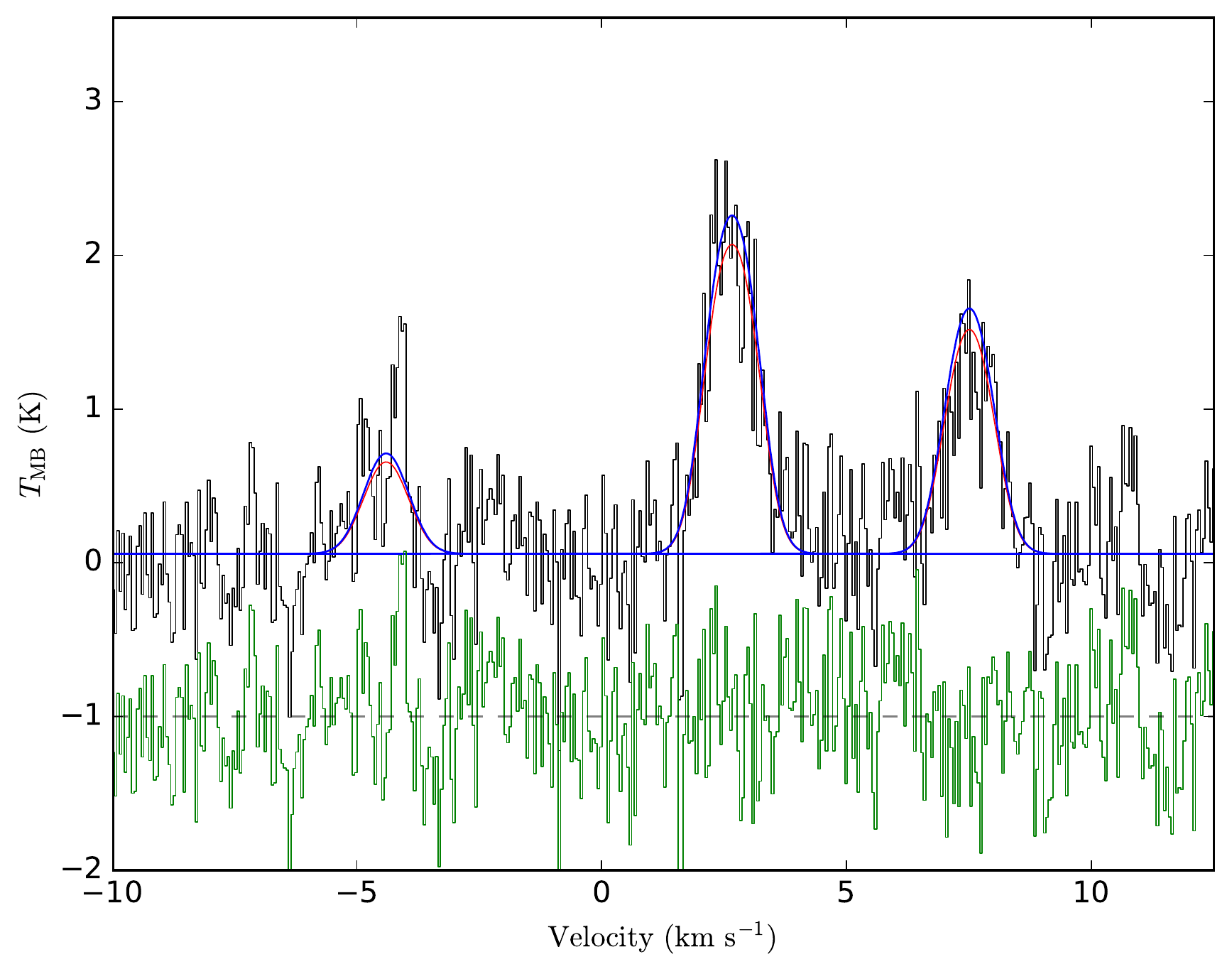}} 
\caption{A hyperfine fit for the HCN spectra from clump B. The blue line is the result of hyperfine fitting and the red line is a simple Gaussian fit to the spectra. The green curve shows the residuals of the fitting. The residuals are vertically offset by -1 K .}
\label{fig:HCN}
\end{figure}

The HCN hyperfine structure can be used to derive additional information about the clump environments. We have used the pyspeckit package \citep{pyspeckit} for fitting the hyperfine structure of all six extracted spectra of HCN, from clumps A-F in Fig. \ref{fig:location}. However the S/N of the HCN lines was too low for all of the clumps, except for clump B (shown in Fig. \ref{fig:HCN}), to produce a reasonable fit.

\section{Additional $^{12} \rm CO$ spectra}

\begin{figure}[!h]
\centering
\resizebox{\hsize}{!}{\includegraphics[width=17cm]{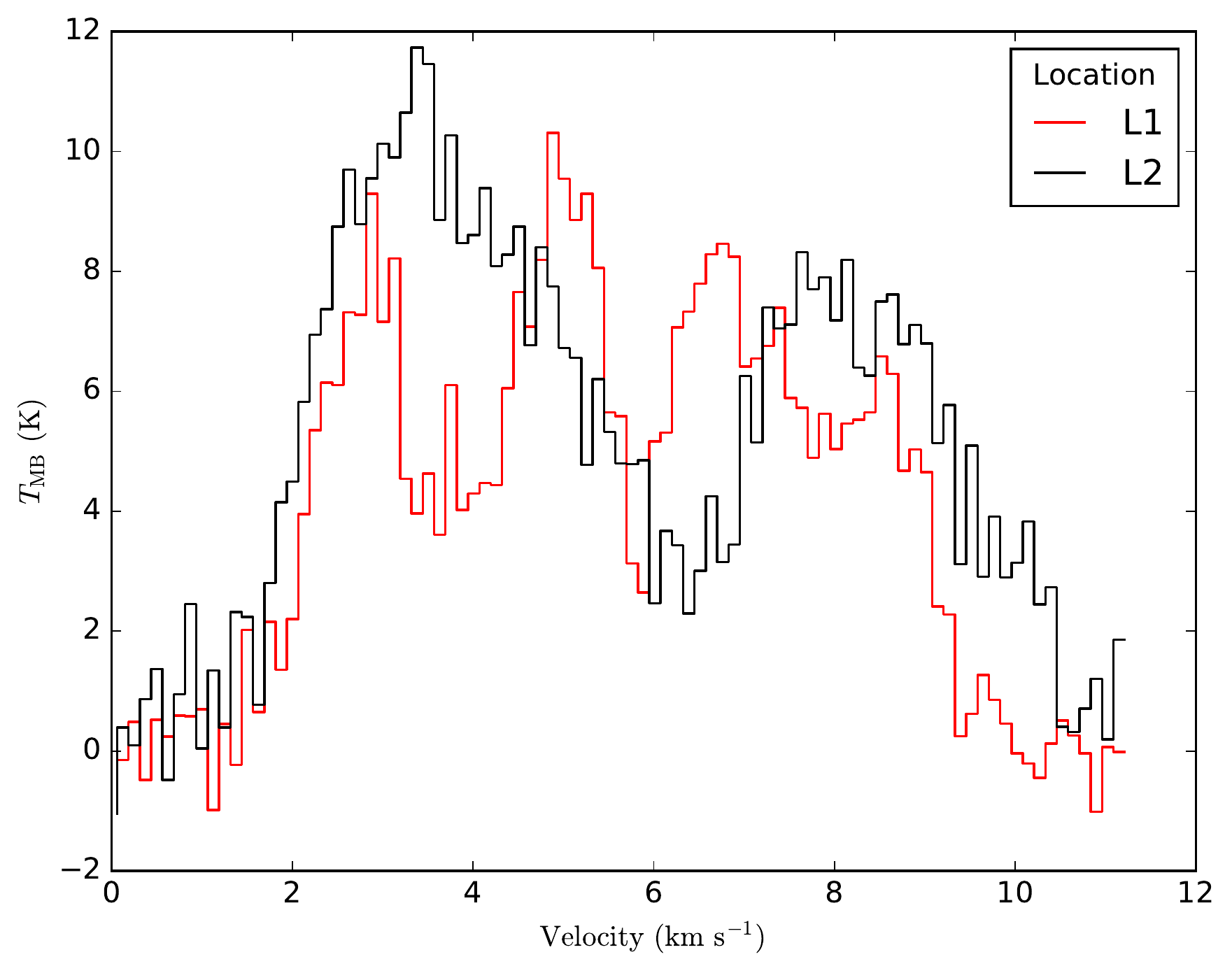}} 
\caption{ $^{12} \rm CO$ ($J=1-0$) spectra extracted from positions L1 and L2 in Fig. \ref{fig:location}.}
\label{fig:third_12CO}
\end{figure}

Shown in Fig. \ref{fig:third_12CO} are the additional Nobeyama $^{12} \rm CO$ spectra extracted from positions L1 and L2 (see Fig. \ref{fig:location}).

\end{appendix}
\end{document}